\documentclass{article}
\usepackage{graphics}
\usepackage{natbib}

\begin{document}

\title{A Graph Model for the Evolution of Specificity in Humoral Immunity}

\author{Luis E. Flores\\
Eduardo J. Aguilar\\
Valmir C. Barbosa\thanks{Corresponding author ({\tt valmir@cos.ufrj.br}).}\\
Lu\'\i s Alfredo V. de Carvalho\\
\\
Universidade Federal do Rio de Janeiro\\
Programa de Engenharia de Sistemas e Computa\c c\~ao, COPPE\\
Caixa Postal 68511\\
21941-972 Rio de Janeiro - RJ, Brazil}

\maketitle

\begin{abstract}
The immune system protects the body against health-threatening entities, known
as antigens, through very complex interactions involving the antigens and the
system's own entities. One remarkable feature resulting from such interactions
is the immune system's ability to improve its capability to fight antigens
commonly found in the individual's environment. This adaptation process is
called the evolution of specificity. In this paper, we introduce a new
mathematical model for the evolution of specificity in humoral immunity, based
on Jerne's functional, or idiotypic, network. The evolution of specificity is
modeled as the dynamic updating of connection weights in a graph whose nodes
are related to the network's idiotypes. At the core of this weight-updating
mechanism are the increase in specificity caused by clonal selection and the
decrease in specificity due to the insertion of uncorrelated idiotypes by the
bone marrow. As we demonstrate through numerous computer experiments, for
appropriate choices of parameters the new model correctly reproduces, in
qualitative terms, several immune functions.

\bigskip
\noindent
{\bf Keywords:} Immune-system specificity, Functional network, Idiotypic
network.
\end{abstract}

\section{Introduction}\label{intro}

The immune system is one of the body's major regulatory systems. One of its
main known functions is to fight agents that are potentially harmful to the
body, including foreign agents and body cells whose behavior is abnormal or
dangerous, as in the case of cancerous or virus-infected cells. These and other
immune functions arise from complex interactions involving numerous molecules
and cells, as well as some of the body's organs. The immunity an individual is
born with is the \emph{innate immunity}. It is highly nonspecific, in the sense
that the mechanisms associated with it are not the result of adaptation during
previous encounters with extraneous agents, but is nonetheless capable of
destroying several types of pathogens. The individual's \emph{acquired
immunity}, on the other hand, is the result of the continual exposition of the
body to the action of extraneous substances, called \emph{antigens}, and tends
to become more specific at each new encounter with the same antigen.

Of the several players involved in acquired immunity, the molecules known
as \emph{cytokines} and \emph{antibodies}, and the cells known as
\emph{B cells}, \emph{helper T cells}, and \emph{cytotoxic T cells}, suffice for
a description of the basic mechanism at a very high level of
abstraction.\footnote{We provide very little detail on the functioning of the
immune system in this paper. The reader is referred to one of the several
textbooks available, as for example \citet{a03}.} When a B cell recognizes an
antigen with which its receptors have high affinity, the cell becomes stimulated
and eventually displays on its surface portions of the antigen. This is one of
the necessary signals for helper T cells to become activated and liberate
cytokines that, in turn, signal the previously stimulated B cells to proliferate
in a process that leads to the production of antibodies that can bind to the
antigen and lead to its destruction. Such a mechanism is the essence of the
so-called \emph{humoral immunity}, the one that takes place in the body's fluids
(or humors) and is mediated by antibodies. The other type of acquired immunity,
known as \emph{cellular immunity}, is also triggered by the cytokines that the
activated helper T cells liberate, and culminates in the destruction of the
cells displaying antigen portions on their surfaces by the cytotoxic T cells.

This basic mechanism of antigen detection and destruction lies at the core of
acquired immunity, but several higher functions of the immune system are known
to take place that need to be accounted for on more solid theoretical
underpinnings. Two notorious such functions are the immune memory and the
ability of the system to discriminate between self and nonself entities. The
leading theoretical framework to explain these and other phenomena is the
\emph{clonal selection theory} \citep{b57,b59,f95}: groups of B cells with
similar recognition capabilities, or \emph{clones}, are selected for
proliferation. According to this theory, some of the B cells that result from
the proliferation elicited by the antigen become memory cells,\footnote{The
question of memory-cell persistence in the absence of the stimulating antigen
remained open for quite some time, but seems to have been settled recently
\citep{mlr00}.} which in turn fight that same antigen more effectively when it
is next encountered. As for the proper discrimination between self and nonself
entities, the current best candidate explanations seem to come from the
``danger theory'' discussed by \citet{bckfmh98}, \citet{rdrm98}, and
\citet{stvdvom98}, which postulates the need for more specific signals for
T-cell activation.

The clonal selection theory is philosophically reductionist, meaning that the
explanations for more and more phenomena are expected to come from discovering
more and more details on how the several molecules and cells involved in
acquired immunity interact. This inherent bias may have caused several important
properties of the immune system to be discovered belatedly, as for example the
involvement of the immune system in several phenomena related to morphogenesis
\citep{g92}. Together with the theory's having so far failed to account for
various other immune-related phenomena, particularly those that bear on
autoimmunity, this bias has resulted in considerable criticism (although,
arguably, some of it appears misdirected \citep{s02}).

Another major theoretical framework in modern immunology is the
\emph{functional} (or \emph{idiotypic}) \emph{network theory} \citep{j74}. This
theory arose in an attempt to address several questions that the clonal
selection theory, being centered on the antigen, seemed unable to answer. For
example, how is the B-cell repertoire regulated before antigens are ever
encountered? Departing from experimental evidence that B cells and T cells
interact with one another in much the same way as they interact with antigens,
the functional network theory postulates that such interactions lead to a
self-organized system out of which immune functions like the immune memory and
the self-nonself discrimination ability emerge naturally.

The functional network theory was met with enthusiasm originally, but interest
in it has waned considerably of late. The reasons for this include the
difficulty of verifying the theory's usefulness in practice and also the fact
that it too, like the clonal selection theory, remained centered on the antigen,
thereby weakening the interest in it as an opposing theory. But the functional
network theory continues to attract the interest of those who recognize the
aesthetic appeal of its elegant systemic approach and that of other similar
autonomous systems \citep{sc01,dct02}. Also, it seems that, of the two
theoretical frameworks, this is the most promising one in terms of where insight
into autoimmunity is expected to come from. Coupled with recent studies on the
appearance of immunity in organisms that never had contact with antigens, these
observations are helping restore the functional network theory to a place of
great relevance in theoretical immunology \citep{c95}.

At various levels of abstraction, and incorporating the postulates of both
clonal selection and the functional network, several proposals have been put
forward of how to model the functioning of the immune system mathematically
\citep{po79,db88,dbsp92,bz97,pw97,hi00,ks00}. In this paper, we introduce a new
model of the functional network. In our model, the network is represented by a
weighted directed graph whose weights correspond to the degrees of affinity
involved in humoral immunity, especially those related to B cells and antigens.
Our model is built on the B model of \citet{db88} and \citet{dbsp92}, and
contributes a new concept in immune-network modeling, namely the evolution of
specificity by the dynamic updating of the graph's weights. What we have found
through several different computational experiments on this model is that it is
capable of reproducing, in qualitative terms, several of the main immune-system
attributes, including the response to antigens, the immune memory, and some
degree of self-nonself differentiation.\footnote{An earlier version of the study
in this paper is found in \citet{fb02}.}

The remainder of the paper is organized into five additional sections. We start
in Section~\ref{fnet} with a brief review of the most relevant aspects of the
functional network theory, then proceed to Section~\ref{bmodel} for the relevant
aspects of the B model. Our model is introduced in Section~\ref{model}, and in
Section~\ref{results} we report on our computational experiments. We conclude
with closing remarks in Section~\ref{concl}.

\section{The functional network}\label{fnet}

B-cell receptors are known as \emph{paratopes}.\footnote{T cells have paratopes,
too. However, since in our model (as in its precursor, the B model) T cells are
not taken into account explicitly, we henceforth omit them from our discussion
whenever possible.} The B cells that result from B-cell proliferation have
paratopes that are not exact copies of those of the original cell, but rather
are the result of the high mutation rates of the process known as
\emph{hypermutation} \citep{ks00}. The antigen regions that can be recognized by
the immune system are called \emph{epitopes}. Ultimately, the immune response is
the result of cell activation due to the affinity, as given by the
complementarity of several types of properties, between paratopes and epitopes.

The immune system's repertoire of paratopes is limited, so in order for its
recognition capabilities to be suitably wide-ranging, it has been argued that
some conditions need to be satisfied \citep{s92,pw97}. These are that each
paratope must recognize a small group of slightly different epitopes, that the
repertoire of paratopes must be in the order of at least $10^6$, and that
paratopes must be randomly distributed along the possible range of different
paratopes (itself limited to a maximum of $10^{15}$ \citep{a03}).

The key observation at the core of the functional network theory comes from a
closer look at the structure of paratopes. There are light and heavy chains, and
in each chain constant and variable regions. The variable regions can bind to
antigens and are known to contain sub-regions that can be recognized by other
paratopes. That is, B cells also have epitopes. The group of epitopes of a B
cell is called an \emph{idiotype}, while each idiotypic epitope is known as an
\emph{idiotope}.

\begin{figure}
\centering
\scalebox{1.00}{\includegraphics{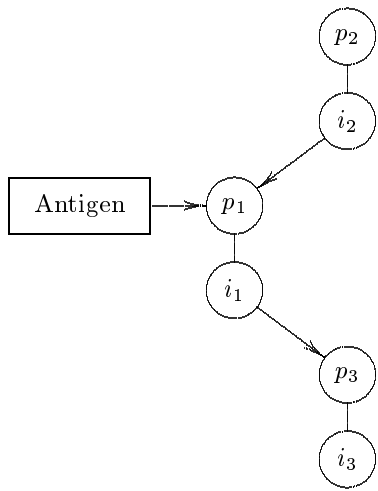}}
\caption{Fragment of a functional network with interfering antigen.}
\label{network}
\vspace{0.45in}
\end{figure}

It follows from this observation that B cells can recognize one another. A
group of B cells having similar paratopes (a clone, as in the clonal selection
theory) is characterized by this set of similar paratopes and also by the
corresponding collective idiotype. These paratope-idiotype pairs, denoted
generically by $p_k$-$i_k$, are illustrated in Figure~\ref{network}. In this
figure, an arrow is drawn from a clone's idiotype to another clone's set of
paratopes to indicate that the former clone stimulates (is recognized by) the
latter. Equivalently, one may think of the arrow as indicating that the latter
clone inhibits (recognizes and seeks to eliminate) the former.

Using Figure~\ref{network} as an example, the functioning of the network can be
intuitively grasped as follows. Before an antigen comes into the body, the
network remains in population equilibrium. Clone $p_1$-$i_1$ stimulates clone
$p_3$-$i_3$, which in turn inhibits clone $p_1$-$i_1$. Meanwhile, clone
$p_2$-$i_2$ stimulates $p_1$-$i_1$ and $p_1$-$i_1$ inhibits $p_2$-$i_2$. This
stimulation-inhibition interplay maintains the clonal population's balance in
the network.

When antigens are introduced in the system and interfere with clone $p_1$-$i_1$,
they cause its population to increase, thus taking the network out of balance.
Both the stimulatory action of $p_1$-$i_1$ over $p_3$-$i_3$ and the inhibitory
action of $p_1$-$i_1$ over $p_2$-$i_2$ increase, which leads the population of
$p_3$-$i_3$ to increase and that of $p_2$-$i_2$ to decrease. As a result,
$p_3$-$i_3$ inhibits $p_1$-$i_1$ more intensely, while $p_2$-$i_2$ stimulates
$p_1$-$i_1$ less intensely. These two forces then concur toward making the
population of $p_1$-$i_1$ decrease, and eventually let it stabilize once again.

\section{The B model}\label{bmodel}

As we indicated earlier, the affinity between paratopes and idiotypes is due 
to the complementarity that exists between molecules in terms of geometric or
physicochemical characteristics. If $c$ is the number of relevant
characteristics, then a $c$-dimensional vector space, known as the \emph{shape
space} \citep{po79}, can be used to formalize the notion of affinity. A point 
$(z_1,\ldots,z_c)$ in the shape space can be taken to represent some
multimolecular structure as far as the $c$ characteristics are concerned. One
possibility to indicate the affinity between this structure and another is to
start by identifying the point in the shape space that corresponds to the
structure that has the greatest affinity with it. If we use symmetry with
respect to $(0,\ldots,0)$, for example, then this point is $(-z_1,\ldots,-z_c)$.

Beyond this characterization of maximum affinity, several other possibilities
exist for the treatment of features of interest when studying the immune system.
For example, suppose we identify a certain clone $C$ with a point $z$ in the
shape space. In order to represent the property that $C$ can be stimulated by
various other clones,\footnote{This is the property known as \emph{cross link}
\citep{pw97}.} not only by the clone $C'$ of maximum affinity to $C$, a sphere
of small radius can be used to represent a set of clones with idiotypes similar
to those of $C'$. This sphere is centered on the point $z'$ of the shape space
that is identified with $C'$. Every point inside the sphere corresponds to a
clone that can stimulate clone $C$ with some significant degree of affinity, but
this degree is ever smaller as the point is located farther from $z'$
\citep{dbsp92}.

One prominent model that employs the shape-space formalism is the B model. This
model has appeared as numerous variations of the same basic idea
\citep{db88,dbsp92,pw97,hi00}, but we limit ourselves to describing only the key
elements on which we build in Section~\ref{model}.

The B model is one of the simplest mathematical representations of the immune
system: it only seeks to describe the dynamics of B-cell populations, and takes
into account the roles of T cells and antibodies only implicitly. For a system
with $n$ clones, the B-model equation for clone $i$, with $1\le i\le n$, is
\begin{equation}
\label{bmodeleq}
\frac{dx_i(t)}{dt}=b+\left[pg\left(h_i(t)\right)-d\right]x_i(t),
\end{equation} 
where $x_i(t)$ is the population of clone $i$ at time $t$, $b$ is the rate at
which new clones are inserted in the system by the bone marrow, $p$ is the rate
of clone proliferation, $d$ is the rate of clone death, $g$ is an activation
function, and $h_i(t)$ is the so-called field of clone $i$ at time $t$.

The field of clone $i$ depends on the affinity between clone $i$ and the other
clones. It is given by
\begin{equation}
h_i(t)=\sum_{j=1}^nw_{ij}x_j(t),
\end{equation}
where $w_{ii}=0$ and $w_{ij}=w_{ji}$ represents the symmetric affinity between
clones $i$ and $j$. This affinity can be determined from neighborhoods in the
shape space, for example, as we indicated earlier.

The activation function $g$ is intended to model the three possible activation
states of a clone: the virgin state, in which the clone is not yet stimulated;
the immune state, in which the clone proliferates; and the inhibited state, in
which proliferation stops. Modeling the inhibited state, in special, is
important because it provides an indirect means of precluding clonal
proliferation due to the overabundance of stimulation, as in the case of
self-antigens, and thereby a means of accounting for the elimination of
self-recognizing paratopes that is known to occur in the immune system.

One possibility for $g$ is to use the window function
\begin{equation}
\label{bmodelact}
g(h)=\left\{
\begin{array}{ll}
1,&\mbox{if $\ell\le h\le u$;}\\
0,&\mbox{otherwise,}
\end{array}
\right.
\end{equation}
where $\ell$ and $u$ delimit the possible values of a clone's field if it is to
remain in the immune state. By (\ref{bmodelact}), the clone remains virgin if
the field is less than $\ell$, it enters the immune state if the field is
between $\ell$ and $u$, or it is in the inhibited state if the field is greater
than $u$.

Some assumptions are implicit in the B model that must be brought to the fore
clearly. One of them concerns the insertion of new clones in the system by the
bone marrow. By adopting the constant rate $b$ for all clones, it has been
assumed that the bone marrow generates B-cell paratopes uniformly randomly
across the entire range of possibilities, independently of what may be taking
place in the immune system (the preferential generation of B cells with certain
paratopes that is known to take place when a specific antigen is being
fought---cf.\ \citet{a03}---is thus ignored). It has been assumed, furthermore,
that this range of possibilities is restricted to the paratopes of the $n$
different clones. Another assumption is that the parameter $p$ suffices to
summarize the role of helper T cells in the proliferation of stimulated B cells
for all clones, and similarly for the parameter $d$ with respect to the
destruction of B cells.

\section{Modeling the dynamics of specificity}\label{model}

Like the B model, our model of humoral immunity concentrates on the dynamics
of B-cell populations and only considers explicitly one of the other major
participants, the bone marrow. It also inherits from the B model the assumptions
discussed in Section~\ref{bmodel} regarding the use of the parameters $b$ and
$p$ to model the functioning of the bone marrow and the proliferation of clones,
respectively.

Notwithstanding these similarities with the B model, the model we introduce in
this section contributes four major innovations to the modeling of the
functional network. The first innovation is the use of clusters of clones, as
opposed to clones, as the basic modeling unit. Each cluster is to be thought of
as a small sphere in the shape space containing clones that, to some degree,
have paratopes (and hence idiotypes) with similar geometric and physicochemical
properties. The adoption of clusters makes the implicit assumption represented
by the $b$ parameter less stringent, since now new paratopes generated by the
bone marrow are assumed to fall within one of the existing clusters, not clones.
We denote by $N$ the total number of clusters.

The second innovation is the explicit use of a directed graph on $N$ nodes (one
for each cluster) to represent the interactions among clusters. A similar graph
has been implicit since Section~\ref{fnet}, for example in Figure~\ref{network}
and also in the underlying connections among clones---via the $w_{ij}$'s---in
the B model. The explicit adoption of the directed graph allows any connection
pattern on the $N$ nodes to be explored, and indirectly a multitude of possible
complementarity criteria among the clones of different clusters.

We denote our directed graph by $D$. In $D$, an edge exists directed from
cluster $i$ to cluster $j$ with $i\neq j$ to indicate that the clones in cluster
$i$ have the capability of stimulating (being recognized by) those in cluster
$j$. Associated with this edge is a positive weight, $w^+_{ij}$, indicating the
average affinity with which this stimulation occurs. The same edge can, of
course, be examined from a dual perspective: clones in cluster $j$ have the
capability of inhibiting (recognizing and seeking to eliminate) those in cluster
$i$. This alternative view leads to another edge weight, $w^-_{ji}$, indicating
the average affinity with which the inhibition occurs. Of course, stimulation
and inhibition between the two clusters occur with the same degree of affinity,
so we have $w^+_{ij}=w^-_{ji}$.\footnote{Note that this is not the same symmetry
that is assumed in the B model. Using the terminology of this section, that
symmetry reads $w^+_{ij}=w^+_{ji}$ or, equivalently, $w^-_{ij}=w^-_{ji}$. We do
not make this assumption.}

The third innovation that our model contributes is a more detailed treatment of
B-cell removal from the system. While in the B model this is accounted for
exclusively by the $d$ parameter, we make a distinction between B-cell removal
by inhibition by other clones\footnote{These are the anti-idiotypic clones of
\citet{j74}.} and removal by other causes (e.g., by apoptosis). The former will
be accounted for by a new term introduced in our model's equations, while the
latter will continue to be modeled by the $d$ parameter.

The last innovation we wish to highlight has also been the major motivation in
this study, and has to do with modeling the evolution of the functional
network's specificity. The idea to be captured by the model is the following.
When cluster $i$ stimulates cluster $j$, those clones in cluster $j$ having
greater affinity with the stimulus tend to proliferate more than the others. As
a result, the distribution of clones in the population of cluster $j$ is altered
and the cluster becomes more specific in terms of its ability to respond to that
stimulus. When the stimulus is withdrawn, the preferential proliferation in
cluster $j$ ceases and the random insertion of new clones by the bone marrow
tends to homogenize the clone population in that cluster once again. In this
case, cluster $j$ becomes less specific with respect to the stimulus.

The notion of stimulus here is intended to capture a quantity that is
proportional to both the idiotypic population in cluster $i$ and the stimulatory
weight of cluster $i$ upon cluster $j$, $w^+_{ij}$. In our model, the
specificity of a cluster, say $j$, is associated with the collective stimulatory
weights of all the edges that in $D$ are directed toward $j$, that is,
$w^+_{ij}$ for every cluster $i$ such that an edge directed from $i$ to $j$
exists in $D$. We allow for the evolution of specificity by letting such weights
be dynamic (and, thus, their inhibitory counterparts as well).\footnote{Another
study of the evolution of specificity on a variation of the B model is the one
of \citet{hi00}, but its authors' approach is altogether different. In
particular, they employ an unchanging matrix of stimulation weights.} This is
done in such a way that the instantaneous change incurred by weight $w^+_{ij}$
is proportional to how close the stimulus provided by cluster $i$ is to being
the greatest individual stimulus received by cluster $j$ at that instant. This
is, clearly, a Hebbian-style rule for the dynamics of specificity. We provide
more details shortly.

Our model's equations are given in terms of a discrete-time parameter $t\ge 0$.
The equation that describes the evolution of the population of clones in cluster
$i$, for $1\le i\le N$, is the extension of (\ref{bmodeleq}) given by
\begin{equation}
\label{modeleq}
x_i(t+1)=x_i(t)+b+\left[pg_n\left(h^+_i(t)\right)-d\right]x_i(t)-qh^-_i(t).
\end{equation} 
In (\ref{modeleq}), $x_i(t)$ is the population of cluster $i$ at time $t$ and
$q$ is the rate of clone removal by inhibition. The field of cluster $i$ at
time $t$ is no longer given as the field of a clone in the B model, but rather
is split into a stimulatory field, $h^+_i(t)$, and an inhibitory field,
$h^-_i(t)$. The activation function is no longer the $g$ of (\ref{bmodelact}),
but rather a ``nodal'' $g_n$---also a zero-one function---to be discussed in
more detail shortly. The remaining entities in (\ref{modeleq}) are all the same
as in the B model.

While most of what is intended with (\ref{modeleq}) is self-explanatory, we
feel the fundamental difference between the addition of new clones by
stimulation and the removal of clones by inhibition requires a little further
elaboration. While, understandably, new clones are inserted in proportion to
how many there are already, the term that accounts for the removal of clones
by inhibition is in (\ref{modeleq}) independent of $x_i(t)$. There are two
reasons for this. The first stems from the fact that the inhibitory potential of
a cluster upon cluster $i$ at time $t$ is the same regardless of the particular
magnitude that the population of cluster $i$ has at that time, since it
ultimately depends on the populations of other clusters. Considering that the
removal of clones from cluster $i$ by inhibition occurs through the binding of
clones in those other clusters to clones in cluster $i$, we see that the number
of clones that get removed by inhibition from time $t$ to time $t+1$ is nearly
independent of the population of cluster $i$ at time $t$. An exception, of
course, is the case of a very low population at time $t$, in which case, to
become completely rigorous, (\ref{modeleq}) should be rewritten to prevent
$x_i(t+1)$ from falling below zero. But we refrain from doing so explicitly for
the sake of clarity.

The second reason is a little more subtle. Consider some hypothetical scenario
in which a sudden populational growth occurs in cluster $i$.\footnote{For
example, while attempting to obtain, in the model, effects similar to those of
passive immunization.} In this scenario, a dependence of the inhibitory term
upon $x_i(t)$ would lead to a sudden increase in the number of clones removed as
well, which in turn could only be explained if a sudden growth were to take
place also in the populations of the clusters exerting the inhibition. This,
however, would only be plausible if we were considering antibody, instead of
B-cell, populations.

Thus, the several rates appearing in (\ref{modeleq}) are to be regarded as being
dimensionally distinct. While $b$ is expressed as a number of clones, $p$ and
$d$ are given as nondimensional quantities. The rate $q$, in turn, is a number
of clones per unit of the inhibitory field.

Before we describe the stimulatory and inhibitory fields of a cluster, we need
additional notation. We let $a_i(t)$ denote the amount of antigen present in
cluster $i$ at time $t$, i.e., antigens with epitopes similar to the idiotypes
of cluster $i$. We also let $s_{ij}(t)$ denote one of the aforementioned
stimuli, specifically the stimulus exerted by cluster $i$ upon cluster $j$ at
time $t$. Clearly, $a_i(t)$ participates in the stimulus of cluster $i$ on
cluster $j$ on equal footing with $x_i(t)$, so we have
\begin{equation}
\label{stimeq}
s_{ij}(t)=w^+_{ij}(t)\left[x_i(t)+a_i(t)\right].
\end{equation}
One should notice, in (\ref{stimeq}), that the time-dependency of the weights is
now accounted for explicitly; this is how weights will be denoted henceforth.
Also, for notational convenience we assume $w^+_{ij}(t)=0$ whenever $i$ and $j$
are such that no directed edge exists from $i$ to $j$, including the case of
$i=j$.

The stimulatory and inhibitory fields of cluster $i$ are given, respectively, by
\begin{eqnarray}
\label{sfield}
h^+_i(t)&=&\sum_{j=1}^Ns_{ji}(t)g_e\left(s_{ji}(t)\right)
\nonumber\\
&=&\sum_{j=1}^Nw^+_{ji}(t)\left[x_j(t)+a_j(t)\right]g_e\left(s_{ji}(t)\right)
\end{eqnarray}
and
\begin{eqnarray}
\label{ifield}
h^-_i(t)&=&\sum_{j=1}^Nw^-_{ji}(t)x_j(t)g_e\left(s_{ij}(t)\right)
\nonumber\\
&=&\sum_{j=1}^Nw^+_{ij}(t)x_j(t)g_e\left(s_{ij}(t)\right).
\end{eqnarray}
The $g_e$ that appears in (\ref{sfield}) and (\ref{ifield}) is yet another
zero-one activation function, this one related to the edges of $D$.

The node- and edge-related activation functions, respectively $g_n$ and $g_e$,
are such that
\begin{equation}
\label{nodeact}
g_n(h)=\left\{
\begin{array}{ll}
1,&\mbox{if $h>0$;}\\
0,&\mbox{otherwise}
\end{array}
\right.
\end{equation}
and
\begin{equation}
\label{edgeact}
g_e(s)=\left\{
\begin{array}{ll}
1,&\mbox{if $\ell\le s\le u$;}\\
0,&\mbox{otherwise.}
\end{array}
\right.
\end{equation}
The $g_n$ in (\ref{nodeact}) is a step function at $h=0$, while the $g_e$ in
(\ref{edgeact}) is the same window function of the B model
(cf.\ (\ref{bmodelact})) with suitably adapted parameter values. Together with
(\ref{sfield}) and (\ref{ifield}), they are meant to elicit the following
behavior from (\ref{modeleq}). The stimulatory field of cluster $i$ is given by
the sum of every stimulus on that cluster that is neither too weak (below
$\ell$) nor too strong (above $u$); if nonzero, this stimulatory field causes
the clones in cluster $i$ to proliferate. Similarly, only clusters that cluster
$i$ stimulates inside the $[\ell,u]$ window contribute to the inhibitory field
of cluster $i$ and thereby to the removal by inhibition of clones from cluster
$i$.

This use of the function $g_e$ can be thought of as a distribution of the role
played by the function $g$ in the B model through the edges that in $D$ are
incoming to cluster $i$. When the windowing mechanism is thus distributed, each
individual stimulus on $i$ must lie within the desired bounds to be effective,
not simply the combined stimuli. This is a more reasonable approach when the
graph's nodes stand for clusters of clones, not simply for clones, and therefore
represent a much greater variety of paratopes and idiotypes. The removal of
clones by inhibition is now also more selective, as we already remarked, and
also the removal of antigens, which for cluster $i$ we model by
\begin{equation}
\label{atgremoval}
a_i(t+1)=a_i(t)-rh^-_i(t),
\end{equation}
where we have refrained from explicitly indicating that $a_i(t+1)$ must be
prevented from falling below zero.

In (\ref{atgremoval}), $r$ is the rate of antigen removal, the same for all
clusters, and is expressed in antigen units per unit of the inhibitory field.
Note that the process described in (\ref{atgremoval}) is entirely analogous to
the process of clone removal by inhibition modeled by the last term of
(\ref{modeleq}). In particular, by the definition of the inhibitory field in
(\ref{ifield}), antigen removal only happens at cluster $i$ if at least one
other cluster is stimulated by cluster $i$ within the corresponding edge's
window. As a by-product, we see that parameter values can be chosen so that any
antigen appearing in overwhelming quantities, like self-antigens, stimulates no
cluster and consequently undergoes no attempts at removal.

We finalize the section by returning to the dynamics of specificity. For
$i\neq j$, our criterion for the update of the stimulatory weight of cluster $i$
on cluster $j$ from time $t$ to time $t+1$ is based on the set $S_j(t)$, which
we define to be the set of clusters whose stimuli on cluster $j$ at time $t$
are, individually, within the $[\ell,u]$ window. Notice that we have placed no
restrictions on $i$ or $j$ beyond their being distinct from each other, which
means that even pairs of clusters not currently connected in a certain direction
in the graph are considered for weight update in that direction.

If $i\notin S_j(t)$, then
\begin{equation}
\label{weightupd0}
w^+_{ij}(t+1)=w^+_{ij}(t)+\rho,
\end{equation}
where $\rho$ is a random variable uniformly distributed in the interval
$[-{\mathrm R}/2,{\mathrm R}/2]$ intended to capture the effect of hypermutation
on inter-cluster affinity. This form of weight update happens, in particular, if
$s_{ij}(t)$ is too large due to the presence of a massive amount of antigen in
cluster $i$, so self-antigens cause no weight changes beyond what is effected
randomly.

On the other hand, if $i\in S_j(t)$, then
\begin{equation}
\label{weightupd1}
w^+_{ij}(t+1)=\left(
\frac{\delta s_{ij}(t)}{\displaystyle\max_{k\in S_j(t)}s_{kj}(t)}
+1-\frac{\delta}{2}
\right)w^+_{ij}(t)+\rho.
\end{equation}
According to (\ref{weightupd1}), the value of $w^+_{ij}(t)$ for $i\in S_j(t)$ is
either expanded or contracted before it is added to $\rho$ to yield
$w^+_{ij}(t+1)$. The factor determining this expansion or contraction has the
Hebbian nature we discussed earlier: within the set $S_j(t)$, the weight
associated with the largest stimulus on cluster $j$ undergoes the greatest
expansion (it is multiplied by $1+\delta/2$); all other weights undergo
proportionally smaller expansions or may even get contracted (in the worst case,
by a factor of $1-\delta(0.5-\gamma)$, where
$\gamma=\ell/\max_{k\in S_j(t)}s_{kj}(t)$). What dictates whether a weight is
expanded or contracted is whether the corresponding stimulus is above or below
half the largest stimulus, respectively.\footnote{An alternative to this half is
to use $\alpha$ such that $0<\alpha<1$, so the expansion/contraction factor in
(\ref{weightupd1}) becomes
$\delta s_{ij}(t)/\max_{k\in S_j(t)}s_{kj}(t)+1-\alpha\delta$. In this case, the
threshold between expansion and contraction occurs as the stimulus $s_{ij}(t)$
becomes less than $\alpha\max_{k\in S_j(t)}s_{kj}(t)$. We use $\alpha=0.5$
throughout, though.} The parameter $\delta$ can be chosen in the interval
$[0,2]$ to regulate the range of possible weight expansion or
contraction.\footnote{The upper limit in this interval can, in principle, be
determined more precisely by finding the value of $\delta$ that solves
$\delta(0.5-\gamma)=1$. But $\gamma$ has a different value for each cluster and
is also time-varying. The upper limit of $2$ comes from setting $\ell$, and
hence $\gamma$, to $0$.}

While (\ref{weightupd0}) and (\ref{weightupd1}) express the essential principle
of our weight-update mechanism, a few observations are in order regarding some
crucial details. First of all, once again complete mathematical rigor requires
both (\ref{weightupd0}) and (\ref{weightupd1}) to be rewritten to prevent
$w^+_{ij}(t+1)$ from becoming negative. In addition, no weight is ever allowed
to grow without bounds, being subjected to an upper bound $W$ representing a
conceptual maximum that the affinity between the clones of any two clusters can
attain. A more precise formulation of our criterion for updating weights is then
the following. Let $w(i,j,t)$ be such that
\begin{equation}
w(i,j,t)=w^+_{ij}(t)+\rho,
\end{equation}
if $i\notin S_j(t)$, or such that
\begin{equation}
w(i,j,t)=\left(
\frac{\delta s_{ij}(t)}{\displaystyle\max_{k\in S_j(t)}s_{kj}(t)}
+1-\frac{\delta}{2}
\right)w^+_{ij}(t)+\rho,
\end{equation}
if $i\in S_j(t)$. Then
\begin{equation}
\label{weightupd2}
w^+_{ij}(t+1)=\left\{
\begin{array}{ll}
0,&\mbox{if $w(i,j,t)<0$;}\\
w(i,j,t),&\mbox{if $0\le w(i,j,t)\le W$;}\\
W,&\mbox{if $w(i,j,t)>W$.}
\end{array}
\right.
\end{equation}

By (\ref{weightupd2}), it is possible for $w^+_{ij}(t+1)$ to become zero. By the
definition of the directed graph $D$, we take this to be synonymous with the
disappearance from $D$ of the edge from $i$ to $j$ between times $t$ and $t+1$.
Similarly, $w^+_{ij}(t+1)$ may also become positive by the addition of the
$\rho$ term in (\ref{weightupd0}), meaning that the edge from $i$ to $j$ may
reappear in $D$ (or appear for the first time, if it never existed) from time
$t$ to time $t+1$. In the sequel, whenever needed we refer to $D$ more precisely
as $D(t)$ to indicate the dynamic character of its edge set.

\section{Computational experiments}\label{results}

We have conducted extensive computational experimentation in order to discover
the properties that emerge out of the model introduced in Section~\ref{model}.
Naturally, the different possibilities for combined parameter values are
overwhelmingly too many, so we have settled for a fixed combination that was
found to elicit qualitatively interesting behavior. The parameter values we have
employed in all the experiments we report henceforth are the ones given in Table
\ref{params}.

\begin{table}
\centering
\caption{Parameter values.}
\begin{tabular}{|c|l|c|}
\hline
Param.&Description&Value\\
\hline
$N$&Number of nodes (clusters) in $D$&$100$\\
$b$&Rate of clone insertion by the bone marrow&$0.05$\\
$p$&Rate of clone proliferation&$0.001$\\
$q$&Rate of clone removal by inhibition&$0.002$\\
$d$&Rate of clone removal by other causes&$0.05$\\
$r$&Rate of antigen removal&$0.05$\\
$\ell$&Least stimulus to contribute to the immune state&$0.1$\\
$u$&Greatest stimulus to contribute to the immune state&$200.0$\\
$\delta$&Range of the weight expansion/contraction factor&$0.05$\\
$\mathrm R$&Range of the random additive term for weight update&$0.05$\\
$W$&Maximum possible weight&$20.0$\\
\hline
\end{tabular}
\label{params}
\vspace{0.45in}
\end{table}

The initial conditions for each experiment were generated randomly by sampling
$x_i(0)$ uniformly from the interval $[0,1]$ for all $i\in\{1,\ldots,100\}$ and
$w^+_{ij}(0)$ from $[0,2]$ for all $i,j\in\{1,\ldots,100\}$ with $i\neq j$.
Notice that the corresponding $100$-node graph $D(0)$ is completely connected
(an edge exists directed from every node to every other node) with probability
$1$, and is therefore structurally meaningless. Even though, for $t>0$, $D(t)$
may evolve into a structurally more interesting graph (we discuss this later in
this section), we often analyze an experiment's outcomes by looking exclusively
at edges whose weights are above a certain minimum. So, in essence, the graphs
we deal with can be regarded as being sparsely connected to the degree that
those threshold weights allow.

For $t\ge 0$, such sparser graphs are all spanning subgraphs of $D(t)$, i.e.,
they have the same set of nodes as $D(t)$. We consider subgraphs of two types.
The first type of subgraph is obtained from $D(t)$ by removing every edge whose
stimulatory weight is not locally significant, in the sense that it is of small
magnitude when compared to the stimulatory weights of the other edges arriving
at the same node. For $f$ such that $0<f<1$, a subgraph of this type is denoted
by $D_f^+(t)$. In $D_f^+(t)$, an edge exists directed from $i$ to $j$ if and
only if $w^+_{ij}(t)\ge fW^+_j(t)>0$, where
\begin{equation}
\label{mainstimulator}
W^+_j(t)=\max_{1\le k\le N}w^+_{kj}(t).
\end{equation}

The second type of spanning subgraph of $D(t)$ that we consider comes from
$D(t)$ by removing those edges whose inhibitory weights are not locally
significant, that is, they are of small magnitude when compared to the
inhibitory weights of the other edges leaving the same node. For $f$ given as
before, we denote a subgraph of this type by $D_f^-(t)$. In $D_f^-(t)$, an edge
exists directed from $i$ to $j$ if and only if $w^-_{ji}\ge fW^-_i(t)>0$, where
\begin{eqnarray}
\label{maininhibitor}
W^-_i(t)&=&\max_{1\le k\le N}w^-_{ki}(t)
\nonumber\\
&=&\max_{1\le k\le N}w^+_{ik}(t).
\end{eqnarray}

\begin{figure}
\centering
\begin{tabular}{c@{\hspace{0.00in}}c}
\scalebox{0.32}{\includegraphics{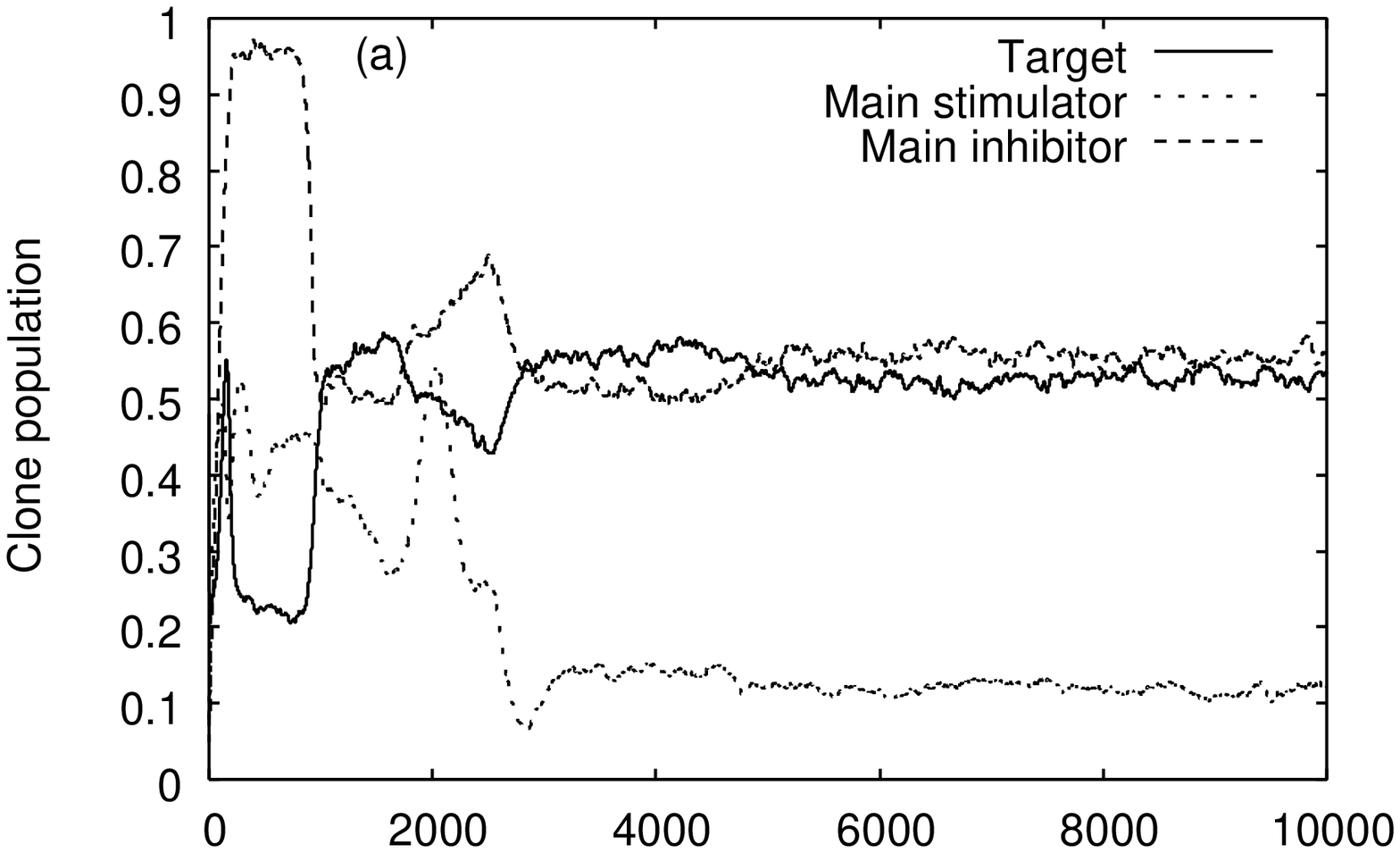}}&
\scalebox{0.32}{\includegraphics{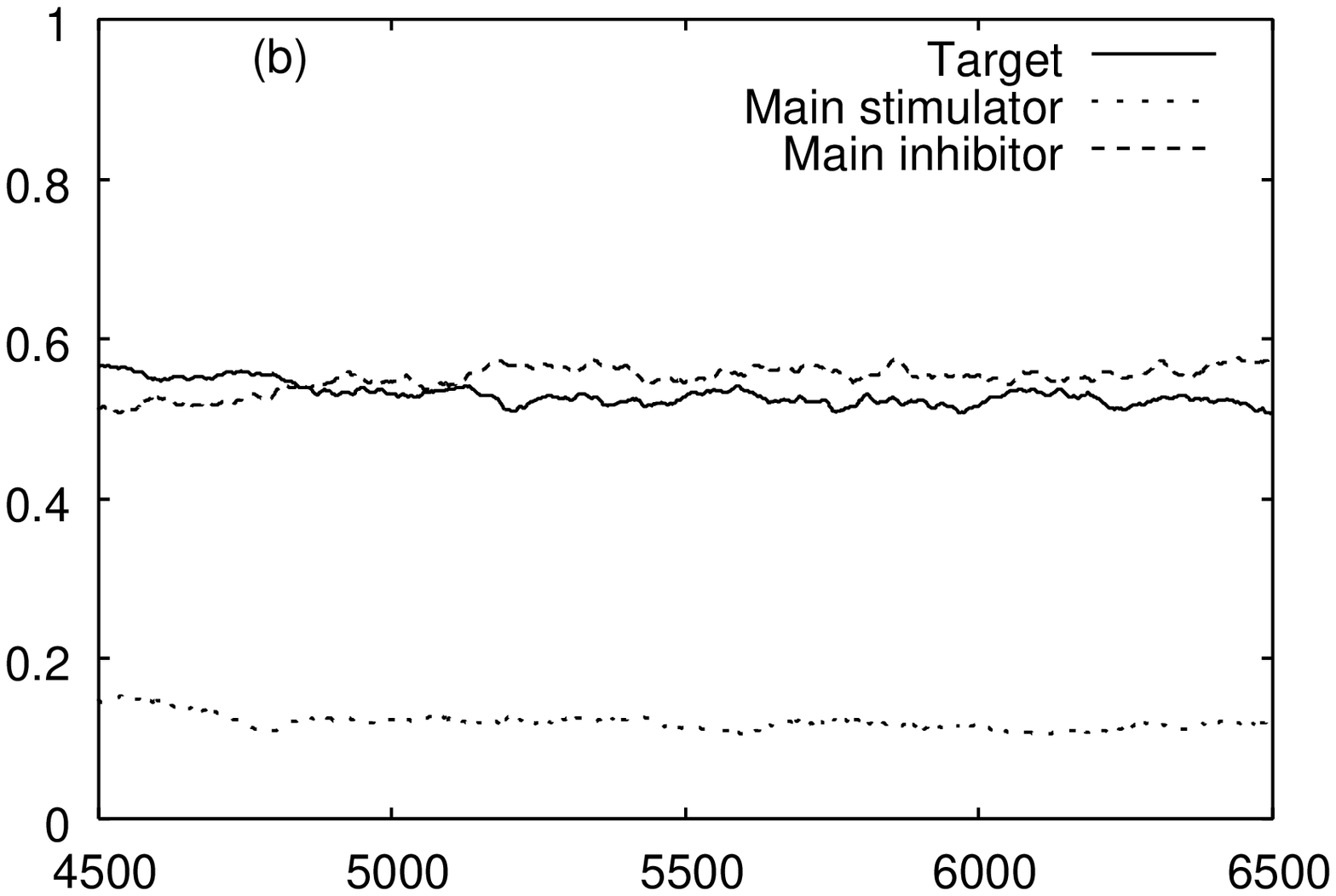}}\\
\scalebox{0.32}{\includegraphics{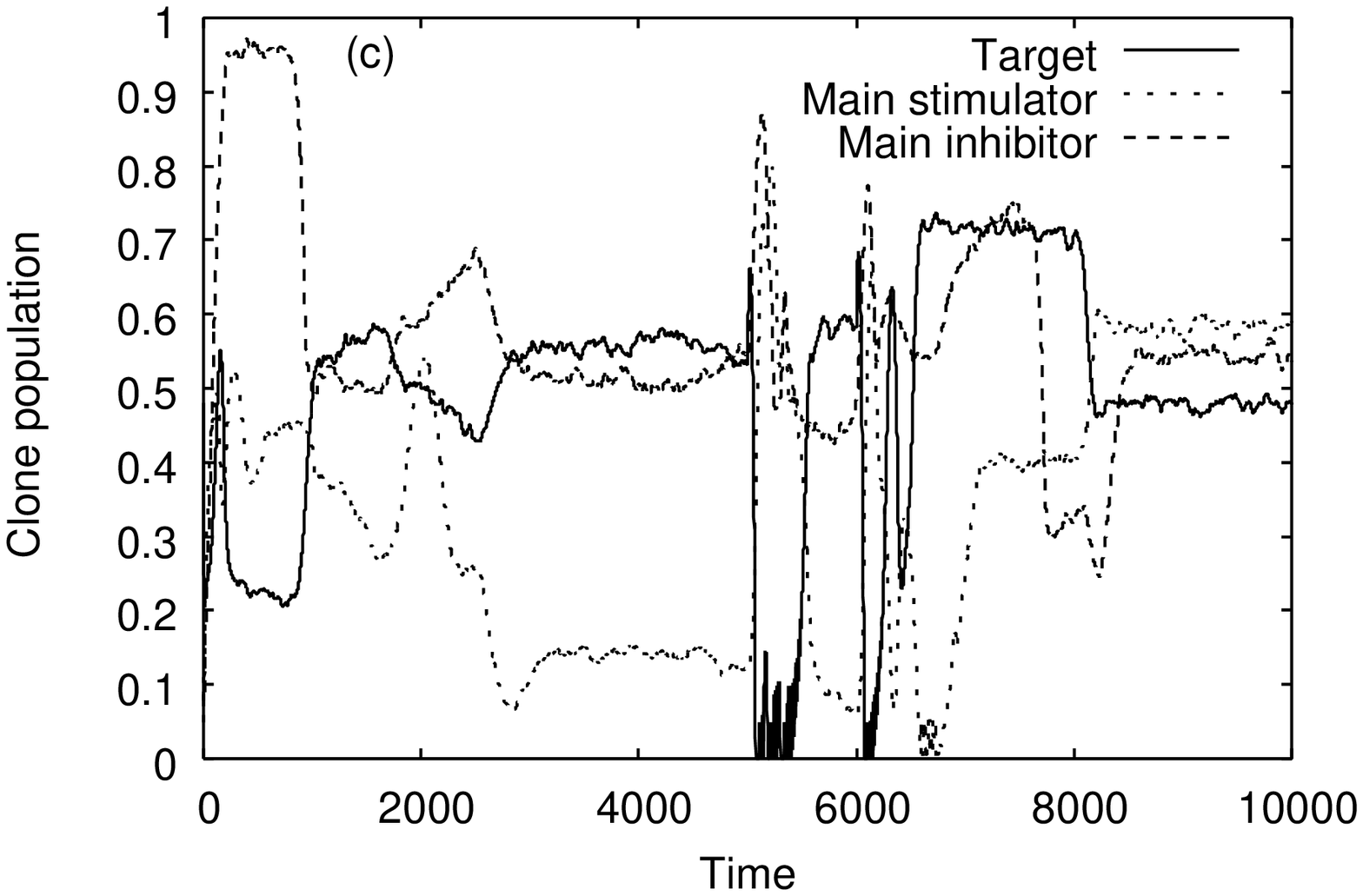}}&
\scalebox{0.32}{\includegraphics{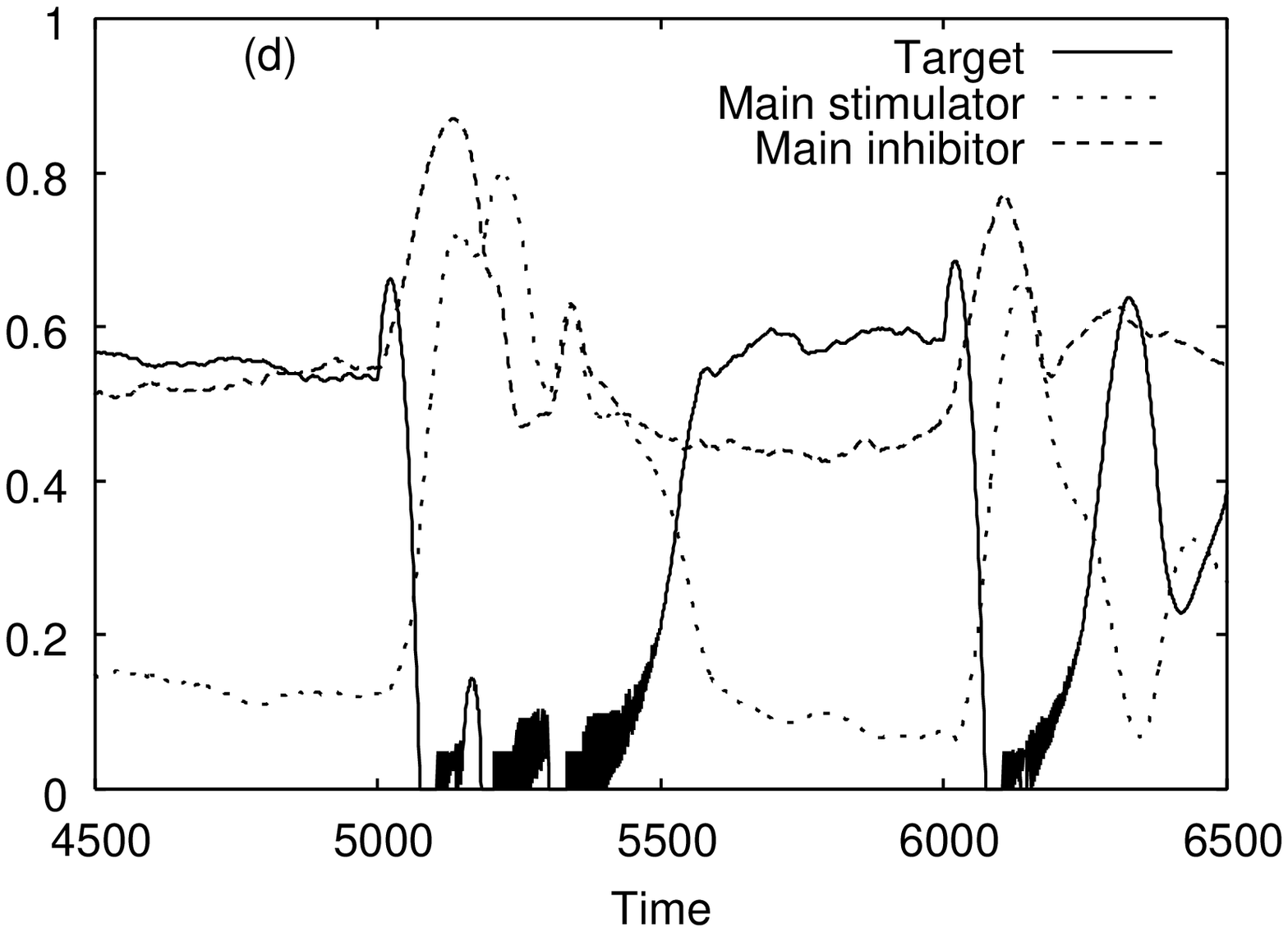}}
\end{tabular}
\caption{Population evolution in the target cluster and in its main stimulator
and main inhibitor. Parts (a) and (b) correspond to no antigen injections. Parts
(c) and (d) correspond to injections of antigen at the target cluster at
$t=5000, 5150, 5300, 6000$ in the amount of $150$ antigen units.}
\label{pop}
\vspace{0.45in}
\end{figure}

We start our presentation of computational results with the data shown in
Figure~\ref{pop}(a), where the evolution of the clone populations of three
clusters is depicted. We found the overall behavior shown in the figure to be
typical across all $100$ clusters, so we selected the three clusters by first
randomly selecting one cluster (the target cluster) and then finding the target
cluster's main stimulator and main inhibitor. The latter clusters are,
respectively, the $k$ that yields the maximum in (\ref{mainstimulator}) when $j$
is the target cluster, and the $k$ that yields the maximum in
(\ref{maininhibitor}) when $i$ is the target cluster.\footnote{Adopting the
denominations of main stimulator and main inhibitor on the basis of the weights
alone may seem confusing, as stimuli depend not only on weights but also on
clone populations. The same holds for an ``inhibition'' that we could define as
each of the terms that make up a cluster's inhibitory field
(cf.\ (\ref{ifield})). We are justified, however, because our weight-update rule
already accounts for adapting weights as a function of the most significant
stimuli. The two denominations are then to be understood as they relate to the
stimulatory and inhibitory weights, respectively.} The typical behavior that is
illustrated in Figure~\ref{pop}(a) is that populations undergo a transient phase
of a few thousand time steps, and then accommodate into a steady regime in which
the oscillations that we expect from \citet{hacf89} and other studies are
nonetheless present. A zoom of Figure~\ref{pop}(a) on two thousand steps of this
steady-regime phase is shown in Figure~\ref{pop}(b).

Figures~\ref{pop}(c) and (d), the latter a zoom of the former on the same two
thousand steps as before, were generated from the same initial conditions that
yielded the data shown in Figures~\ref{pop}(a) and (b), including the seed for
the random-number generator in order to ensure the same evolution. In fact, this
is what happens up until $t=5000$, where we started a series of four antigen
injections at the target cluster. This series consisted of setting $a_i(5000)$,
$a_i(5150)$, $a_i(5300)$, and $a_i(6000)$ to $150$ for $i$ the target cluster.
As shown in Figure~\ref{pop}(c), significant populational disturbances occur in
all three clusters, with a brief attempt at returning to stability shortly
before $t=6000$ and then a return to stability after roughly $t=8000$. Examining
the zoom in Figure~\ref{pop}(d) reveals additional details. Specifically, each
antigen injection elicits a boost in the population of the target's main
inhibitor, probably due to the activation of proliferation in that cluster. The
target's population, in turn, becomes severely depleted, reflecting a
significant increase in its inhibitory field as the antigen is fought. At the
target's main stimulator the population is seen to increase significantly as a
consequence, since inhibition from the target nearly disappears as its
population sinks.

\begin{figure}
\centering
\scalebox{0.50}{\includegraphics{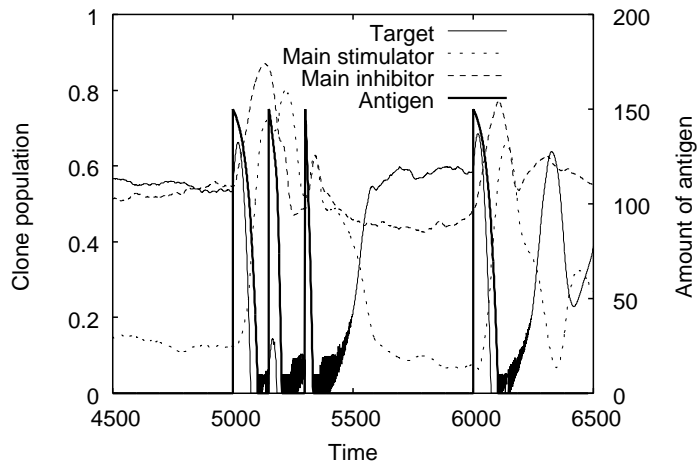}}
\caption{Antigen removal in the scenario of Figure~\ref{pop}.}
\label{ant}
\vspace{0.45in}
\end{figure}

What happens to the antigens at the target cluster in the meantime is shown
in Figure~\ref{ant}, which is essentially a reproduction of Figure~\ref{pop}(d)
with the evolution of the antigen amount in the foreground. Notice that the
antigen is always successfully removed. The first three antigen injections,
occurring relatively in close succession to each other, reveal unequivocally the
presence of the immune memory that the increase in specificity is expected to
entail. The fourth injection, occurring relatively farther in the future, finds
the system with the specificity to that antigen restored to its old situation,
and removal is once again only as fast as when the first injection was
performed.

Our remaining computational results refer to the structure of the graphs $D(t)$,
$D_f^+(t)$, and $D_f^-(t)$, for suitable $f$ and $t$. We analyze the structure
of these graphs by concentrating on their nodes' degrees. The degree of a node
in a graph is the number of nodes to which it is directly connected. In the case
of a directed graph, we split a node's degree into its in-degree (number of
nodes from which a directed edge arrives) and out-degree (number of nodes to
which a directed edge departs).

\begin{figure}
\centering
\begin{tabular}{c@{\hspace{0.00in}}c}
\scalebox{0.32}{\includegraphics{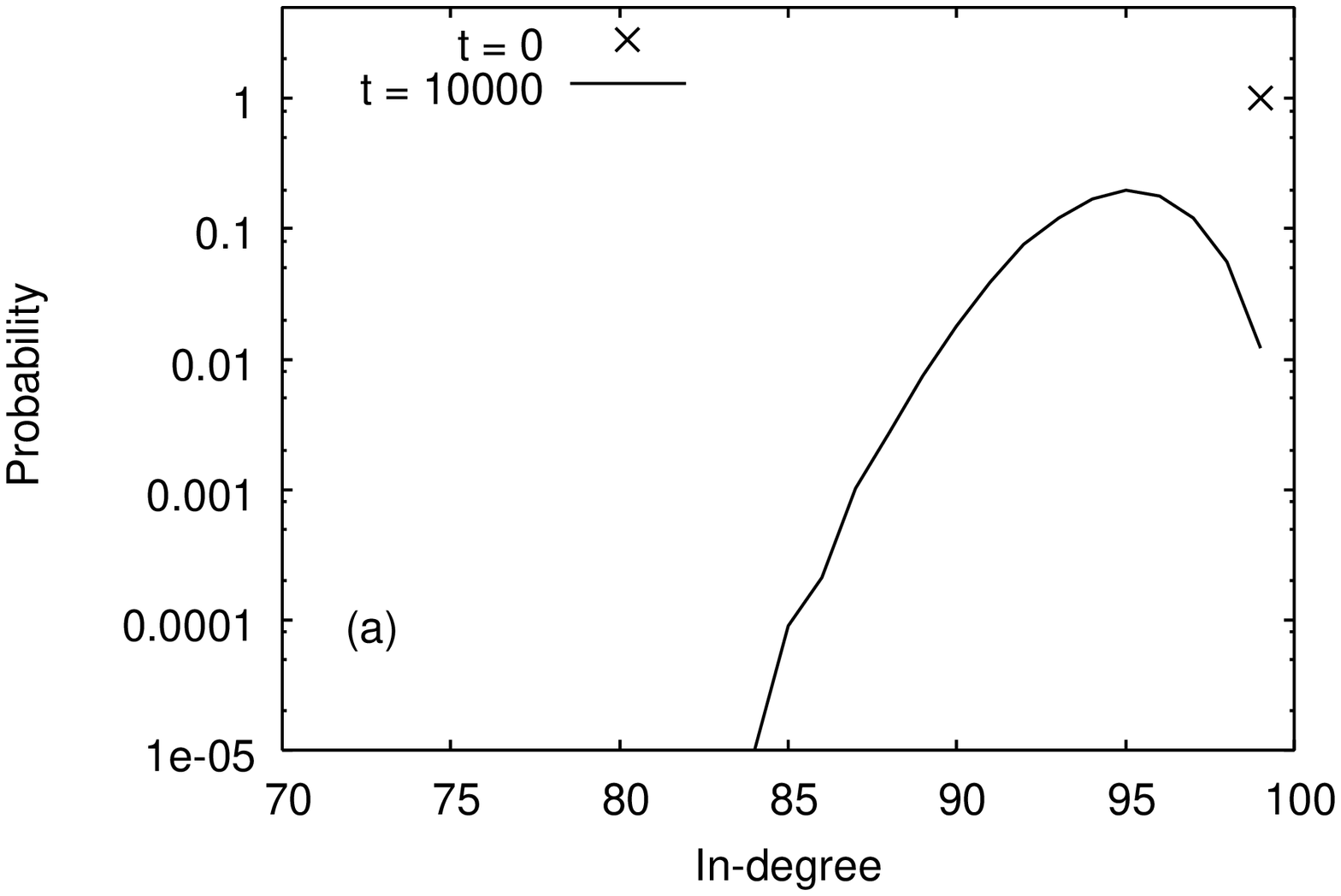}}&
\scalebox{0.32}{\includegraphics{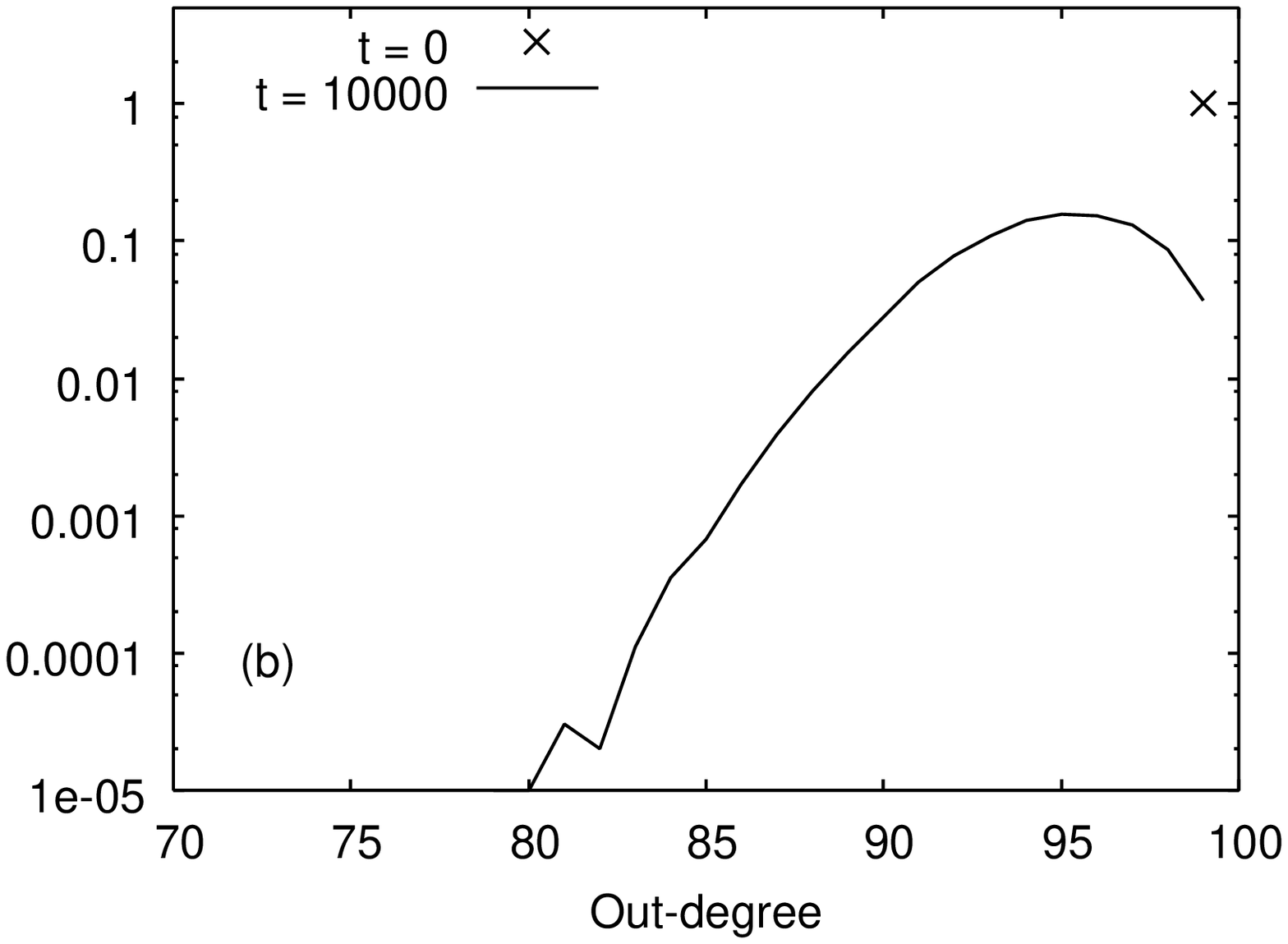}}
\end{tabular}
\caption{Average in-degree distribution (a) and out-degree distribution (b) of
$D(t)$ ($10^3$ independent runs).}
\label{degdist}
\vspace{0.45in}
\end{figure}

We start in Figure~\ref{degdist} by showing the average in- and out-degree
distributions of $D(t)$ as obtained from $10^3$ independent runs. Clearly, as we
anticipated earlier in this section, $D(0)$ is in fact completely connected. In
addition, the passage of time tends to widen the distributions considerably due
to the appearance of zero weights (i.e., the disappearance of edges), but the
distribution continues to be centered around very high means. $D(t)$ is then
very densely connected throughout.

Let us then investigate what happens with the sparser subgraphs of $D(t)$,
starting with their average degrees. Although a node's in- and out-degree may
differ from each other, summing up the in-degrees of all nodes or the out-degree
of all nodes must yield the total number of directed edges in the graph.
Consequently, the graph's average in- and out-degree are the same, so we may
refer to it simply as an average degree.

\begin{figure}
\centering
\begin{tabular}{c@{\hspace{0.00in}}c}
\scalebox{0.32}{\includegraphics{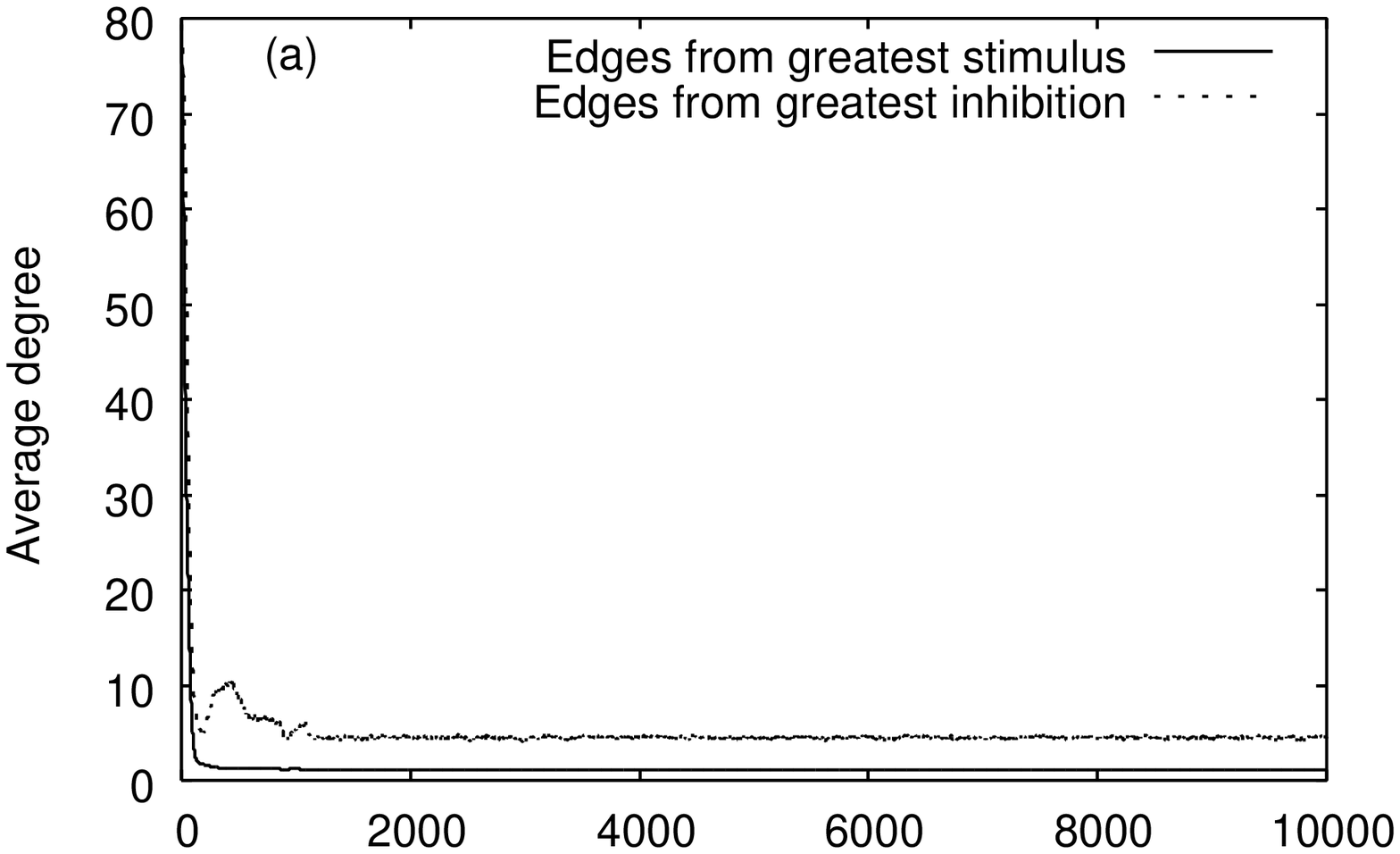}}&
\scalebox{0.32}{\includegraphics{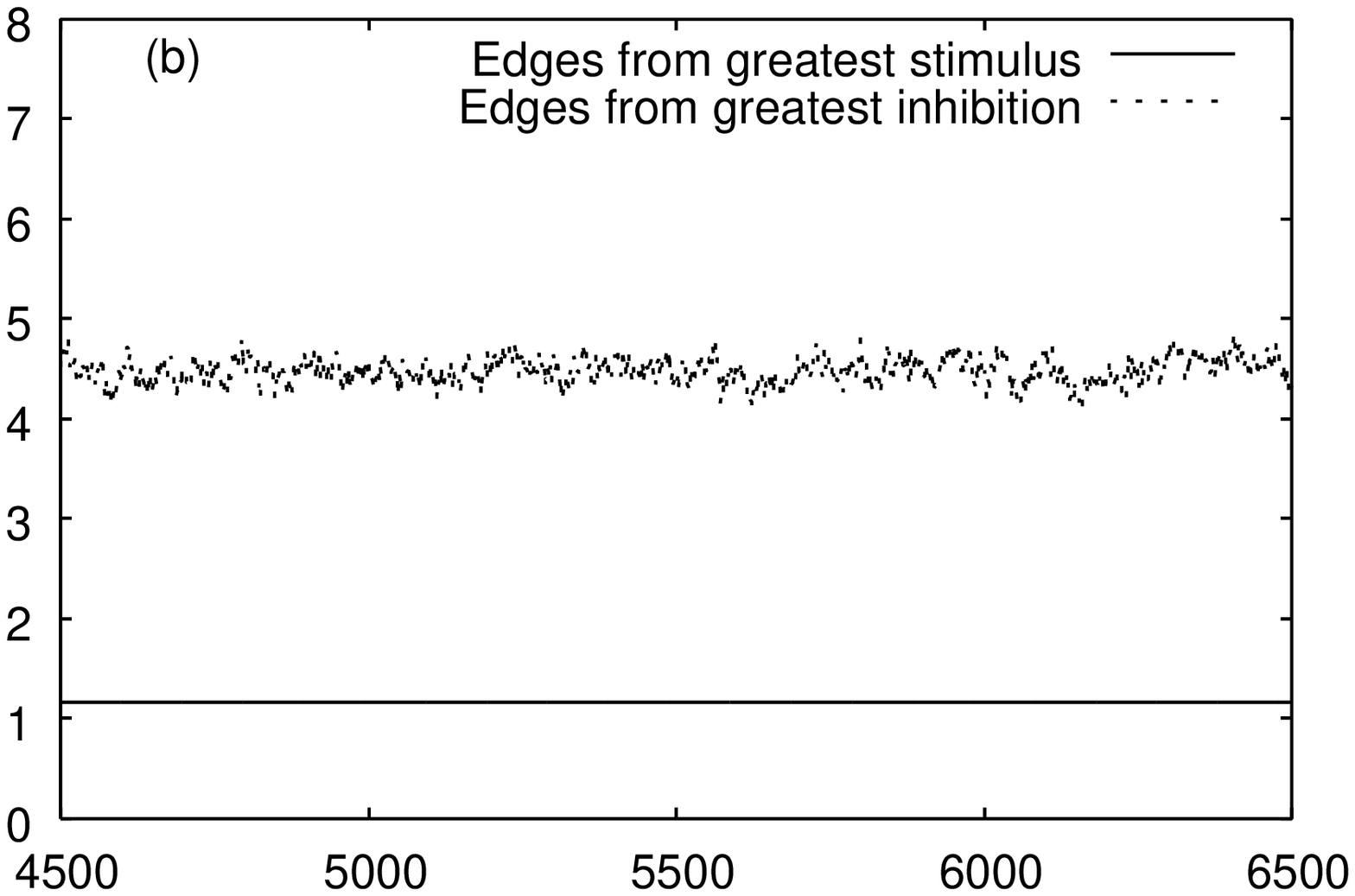}}\\
\scalebox{0.32}{\includegraphics{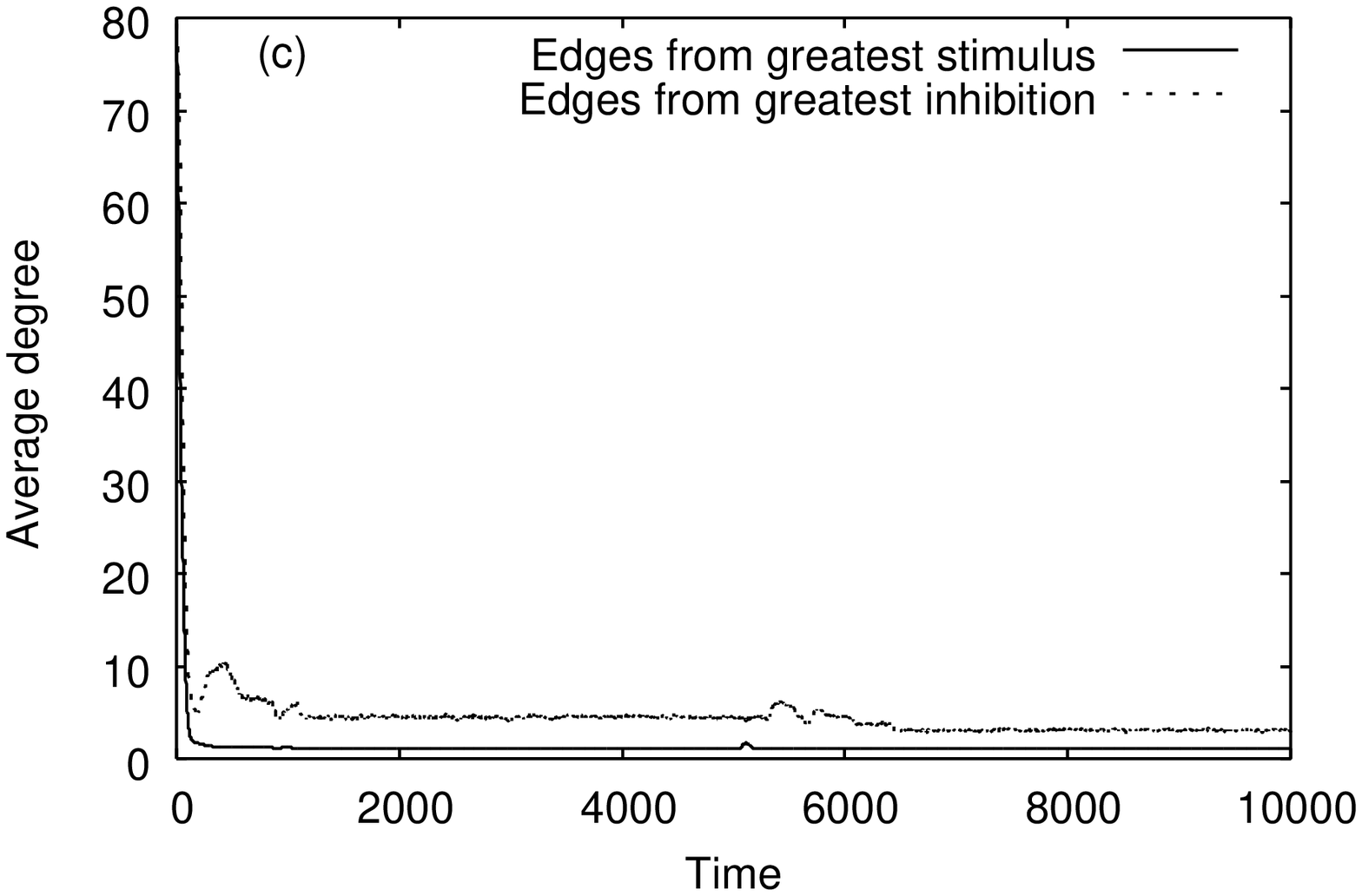}}&
\scalebox{0.32}{\includegraphics{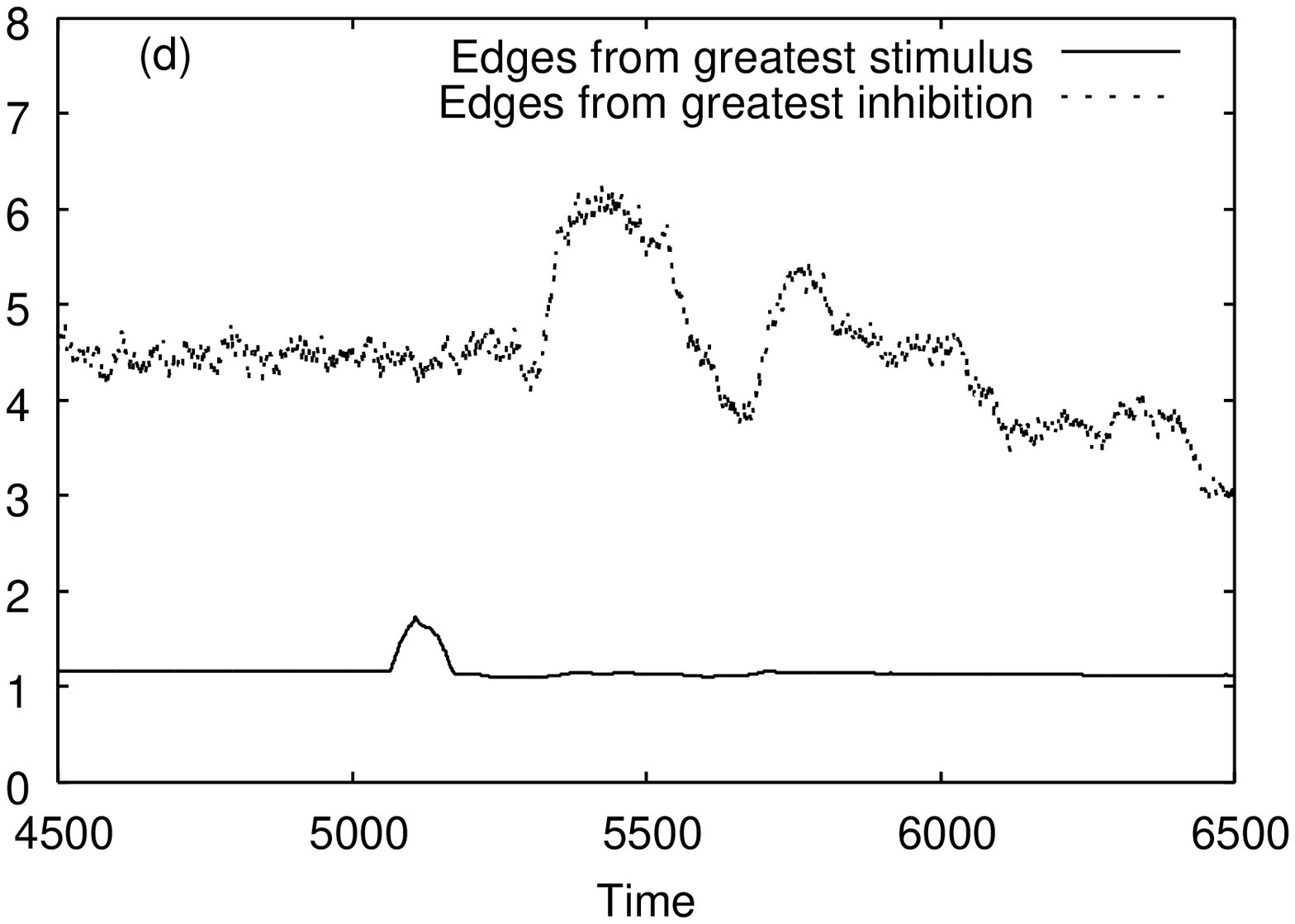}}
\end{tabular}
\caption{Evolution of the average degree of $D_{0.2}^+(t)$ (``Edges from
greatest stimulus'') and $D_{0.2}^-(t)$ (``Edges from greatest inhibition'').
Parts (a) and (b) correspond to no antigen injections. Parts (c) and (d)
correspond to an injection of antigen at the target cluster at $t=5000$ in the
amount of $150$ antigen units.}
\label{avgdeg}
\vspace{0.45in}
\end{figure}

We show in Figure~\ref{avgdeg}(a) the evolution of the average degree of
$D_{0.2}^+(t)$ (labeled ``Edges from greatest stimulus'') and of $D_{0.2}^-(t)$
(labeled ``Edges from greatest inhibition'') in the absence of antigens. Our
choice of $f=0.2$ seeks to privilege edges whose weights at time $t$ are within
a large factor (of $1-f$) of the weights that are locally most significant. The
main effect of this choice is, as we will see shortly, that it yields two
subgraphs of $D(t)$ that differ from each other significantly. Higher values of
$f$ tend to, expectedly, lead to two subgraphs that resemble each other very
closely, in addition to having in- and out-degree more concentrated near one for
practically all nodes if $t$ is large (i.e., way inside the system's
steady-regime zone). What we see in Figure~\ref{avgdeg}(a), of which a zoom is
given in Figure~\ref{avgdeg}(b), is that the average degrees of both graphs
stabilize relatively early in the evolution. Also, in $D_{0.2}^+(t)$ nodes tend
to have in-degree (our out-degree) slightly above one on average, while in
$D_{0.2}^-(t)$ a higher figure is obtained with small oscillations around a
center value.

Repeating the experiment from the exact same initial conditions (including the
seed for random-number generation) yields what is shown in
Figures~\ref{avgdeg}(c) and (d) (the latter a zoom on the former) for an antigen
injection of $a_i(5000)=150$, $i$ the target cluster. Clearly, such a
disturbance affects the graphs' average degrees, briefly in the case of
$D_{0.2}^+(t)$, but in a more lasting fashion for $D_{0.2}^-(t)$.

\begin{figure}
\centering
\begin{tabular}{c@{\hspace{0.00in}}c}
\scalebox{0.32}{\includegraphics{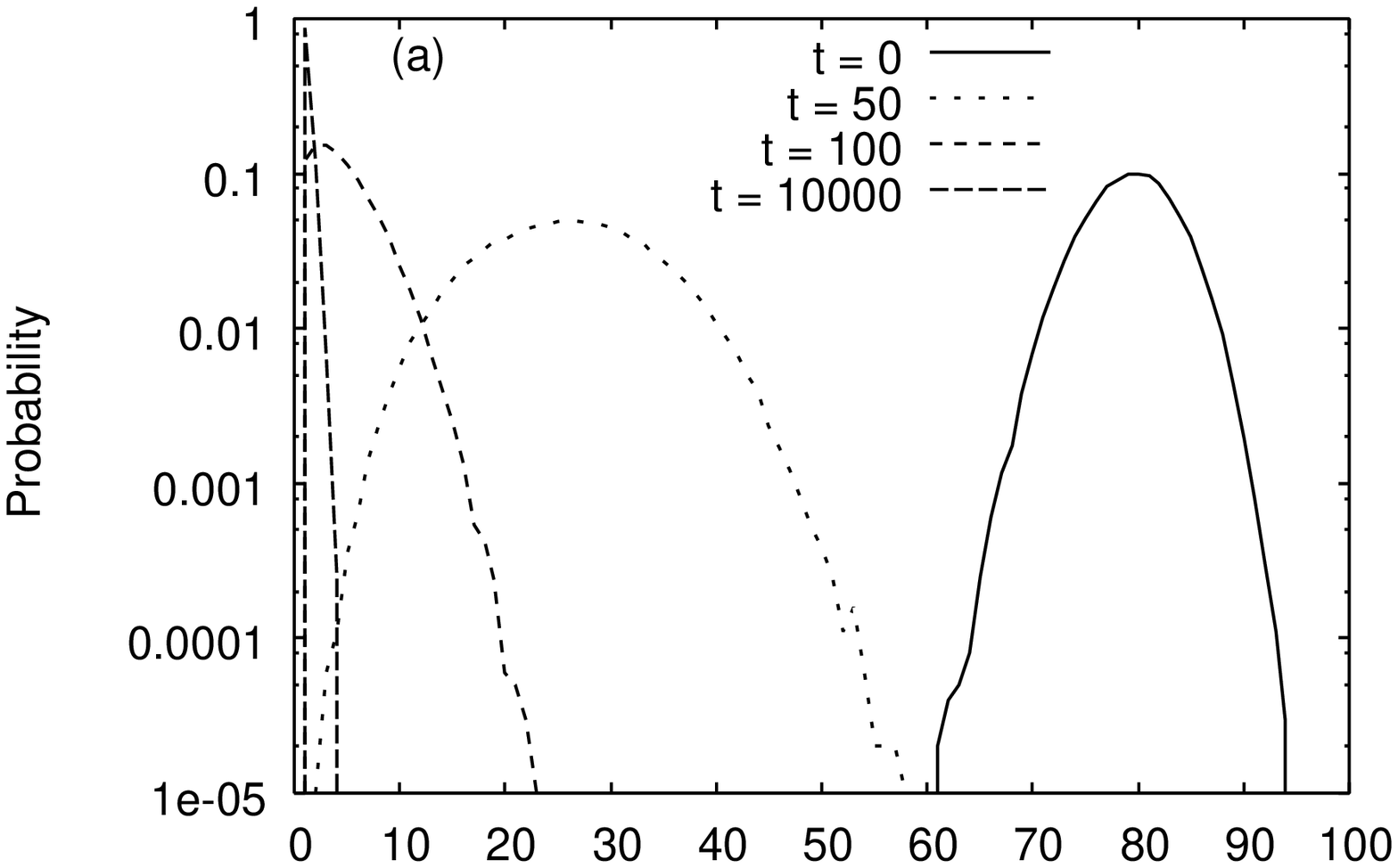}}&
\scalebox{0.32}{\includegraphics{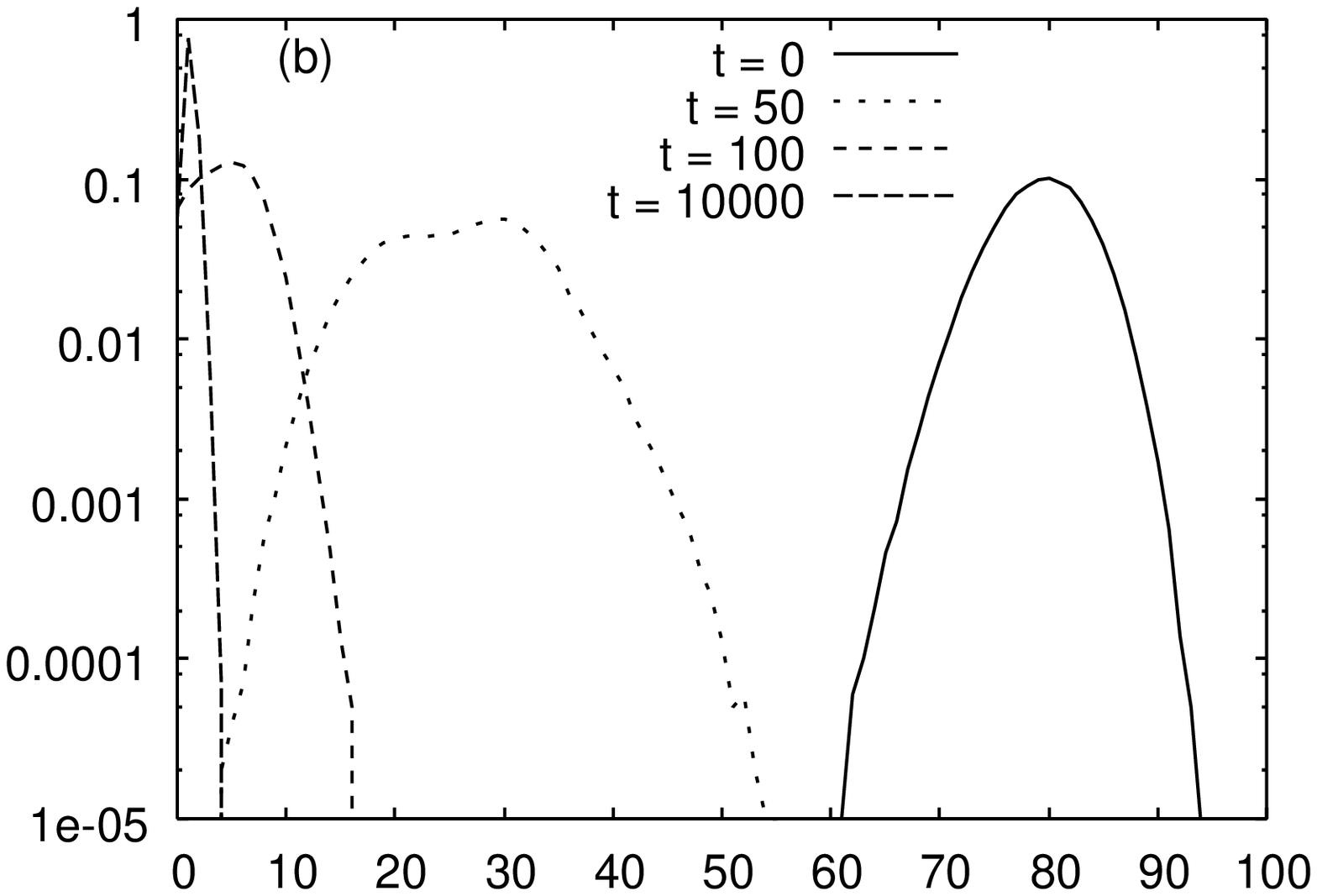}}\\
\scalebox{0.32}{\includegraphics{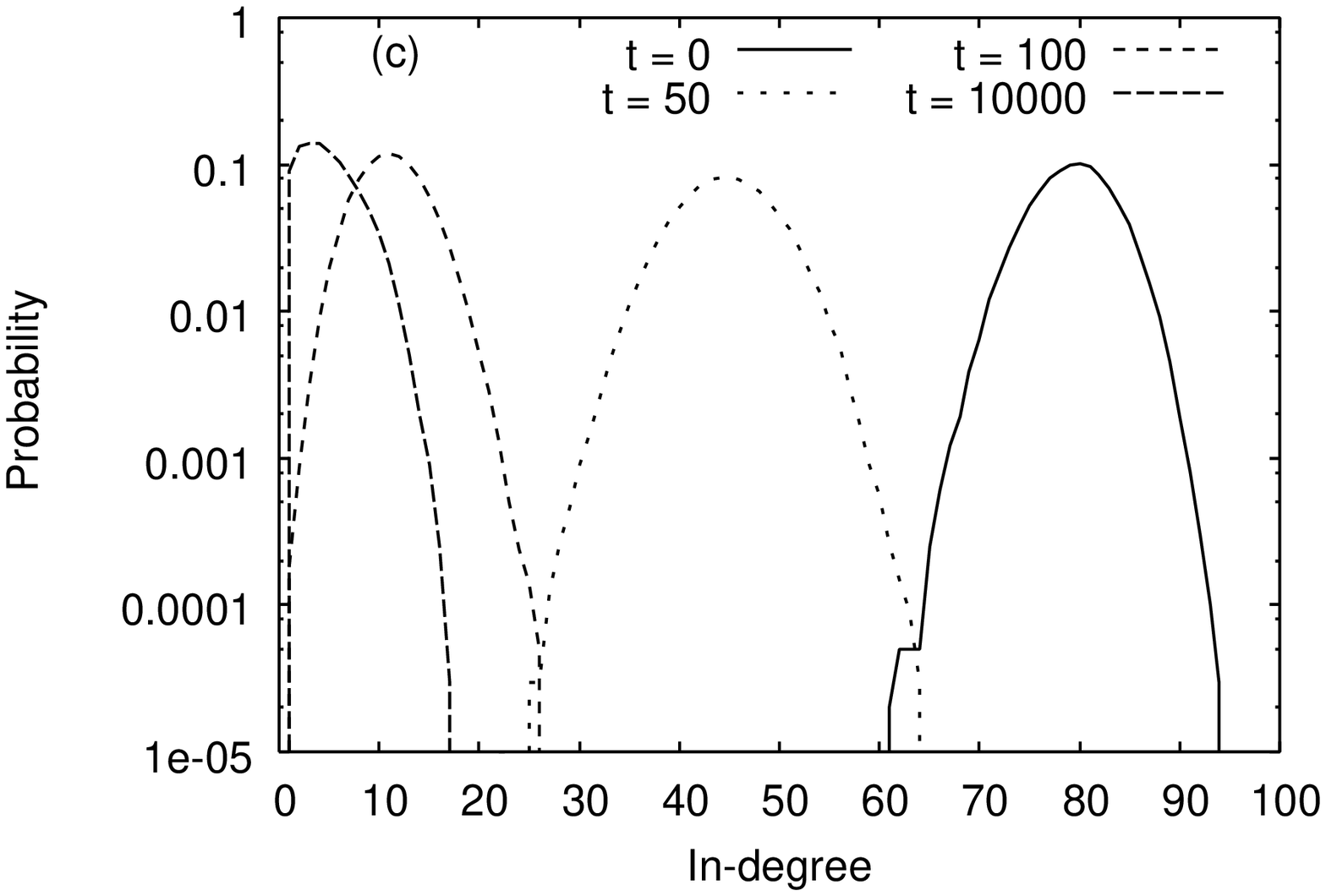}}&
\scalebox{0.32}{\includegraphics{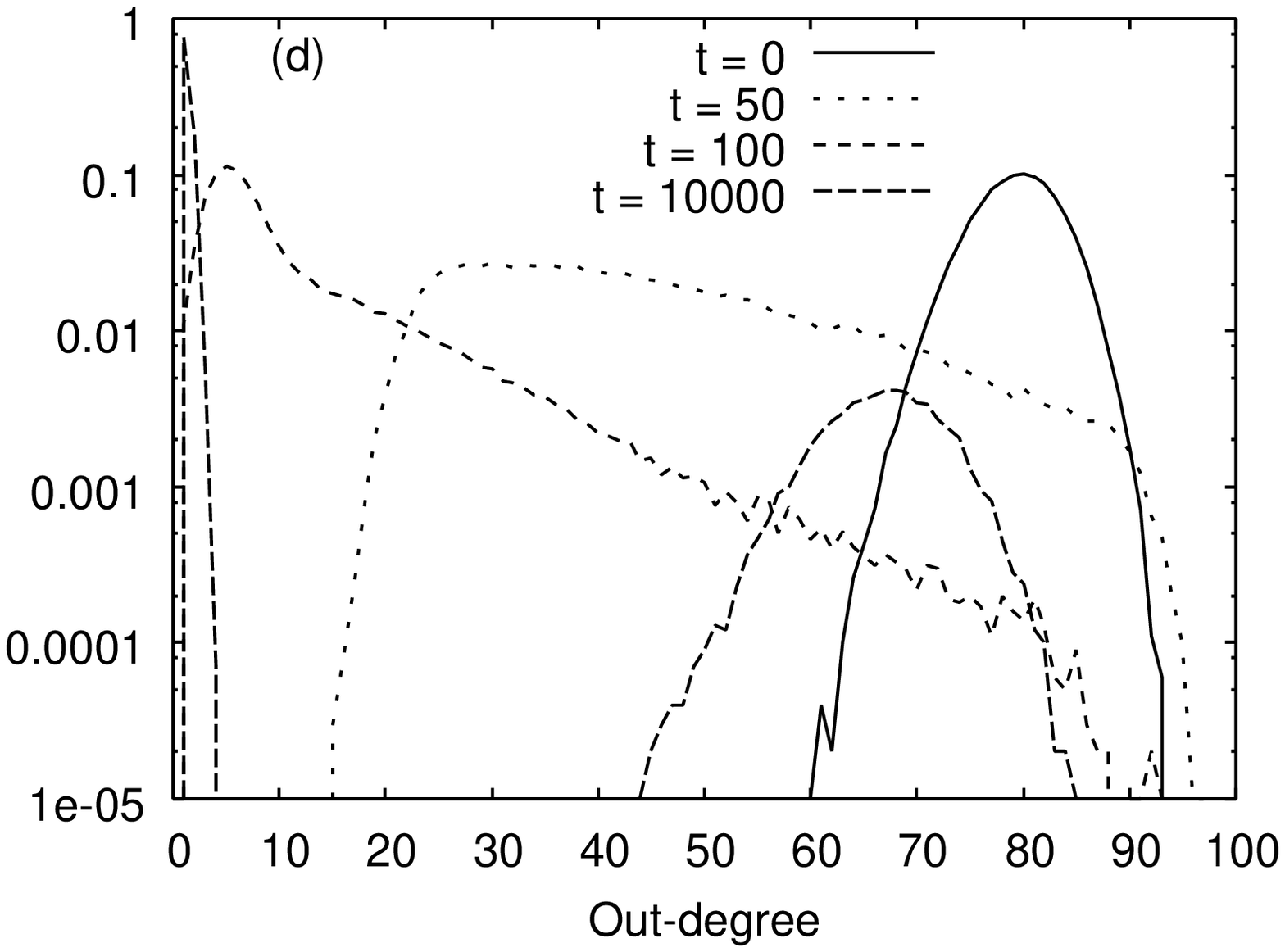}}
\end{tabular}
\caption{Average degree distributions ($10^3$ independent runs). Parts (a) and
(c) give, respectively, the in-degree distributions of $D_{0.2}^+(t)$ and
$D_{0.2}^-(t)$. Parts (b) and (d) give, respectively, the out-degree
distributions of $D_{0.2}^+(t)$ and $D_{0.2}^-(t)$.}
\label{degdist2}
\vspace{0.45in}
\end{figure}

Additional insight into the structure of $D_{0.2}^+(t)$ and $D_{0.2}^-(t)$ can
be gained by looking at how the graphs' degrees are distributed for each value
of $t$. Now, of course, it is once again necessary to separate in- and
out-degrees. Our results are shown in Figure~\ref{degdist2} as the average
outcomes of $10^3$ independent runs. We show the resulting in-degree
distributions of $D_{0.2}^+(t)$ and $D_{0.2}^-(t)$ (parts (a) and (c),
respectively), and the resulting out-degree distributions of $D_{0.2}^+(t)$ and
$D_{0.2}^-(t)$ (parts (b) and (d), respectively). All distributions start off
nearly symmetrically shaped around the value of $80$ (given by $(1-f)N$, as
expected), but very quickly migrate to the relatively tight concentration around
relatively small values we have come to expect from Figure~\ref{avgdeg}.

But Figure~\ref{degdist2} also evidences important differences between the
degree distributions of $D_{0.2}^+(t)$ and those of $D_{0.2}^-(t)$. While the
two distributions of $D_{0.2}^+(t)$ become tightly concentrated around their
means as $t$ grows, in the case of $D_{0.2}^-(t)$ this only seems to hold for
the in-degree distribution, although much more spread than the distributions of
$D_{0.2}^+(t)$. The out-degree distribution of $D_{0.2}^-(t)$ for large $t$ has
the same narrow peak around a very low value, and also a wider but less
pronounced peak at the higher end of the degree range. This wider peak indicates
the occasional occurrence in $D_{0.2}^-(t)$ of nodes with high out-degrees.

These results can be taken back to our original context of modeling the
evolution of specificity and re-interpreted there. When we consider the average
affinities with which the stimulation of clusters upon one another occurs,
and only take into account those affinities that are sufficiently strong with
respect to some thresholding criterion that is local to the stimulated cluster,
then we have seen that in the long run clusters tend to be chiefly stimulated by
very few other clusters, very likely by only one. Conversely, and likewise, a
cluster tends to be the most significant stimulator of very few other clusters,
again most likely only one. This is, of course, consistent with our underlying
approach of increasing stimulatory weights to reflect increases in specificity.

The system can also be examined from a dual perspective, namely the one that
considers the average affinities with which clusters are inhibited by one
another. Such affinities are, of course, the same as the average affinities
with which stimulation occurs. But looking at them from this perspective makes
it possible to filter out, once again by a thresholding mechanism that now is
local to the inhibited cluster, every average affinity that represents an
insignificant inhibition. When we do this, we have seen that in the long run a
cluster can be among the chief inhibitors of a significant number of other
clusters. Conversely, most clusters have one single principal inhibitor,
although there may exist a few clusters for which a very high number of such
inhibitors exist.

\section{Closing remarks}\label{concl}

We have in this paper introduced a new model of the idiotypic network. Our model
has roots in the B model but departs significantly from it by the introduction
of four major innovations: the use of clusters of clones, as opposed to clones,
as the basic modeling unit; the explicit use of a directed graph to describe
affinities and complementarity criteria with great flexibility; a treatment of
B-cell removal that separates removal by inhibition from removal by other
causes; and, most significantly, a framework for modeling the evolution of the
network's specificity by the evolution of edge weights.

Our numerous computational experiments, of which we showed representative
examples, indicate a great ability of the model to capture, in qualitative
terms, some of the major functions of humoral immunity. We find our model to be
quite flexible, and believe it may grow by the incorporation of several
modifications targeted at modeling different specificity-evolution rules, for
example, or the incorporation of more detailed modeling elements aimed at
reflecting current biological knowledge.

But even at its current level of abstraction the model is still challenging in
a number of interesting ways. For example, we would like to detect signs, if
they exist, that our model gives rise to critically self-organized behavior
(cf., e.g., \citet{j98} and \citet{b99}). One of the telltale signs of this
type of behavior in other network models has been the finding that degrees are
distributed according to a power law \citep{br03}, that is, that the
probability that a degree equals $k\ge 0$ is proportional to $k^{-\alpha}$ for
some $\alpha>0$. Our results in Section~\ref{results}, particularly the ones
shown in Figure~\ref{degdist2}, are generally not supportive of the existence of
such a distribution.

\begin{figure}
\centering
\begin{tabular}{c@{\hspace{0.00in}}c}
\scalebox{0.32}{\includegraphics{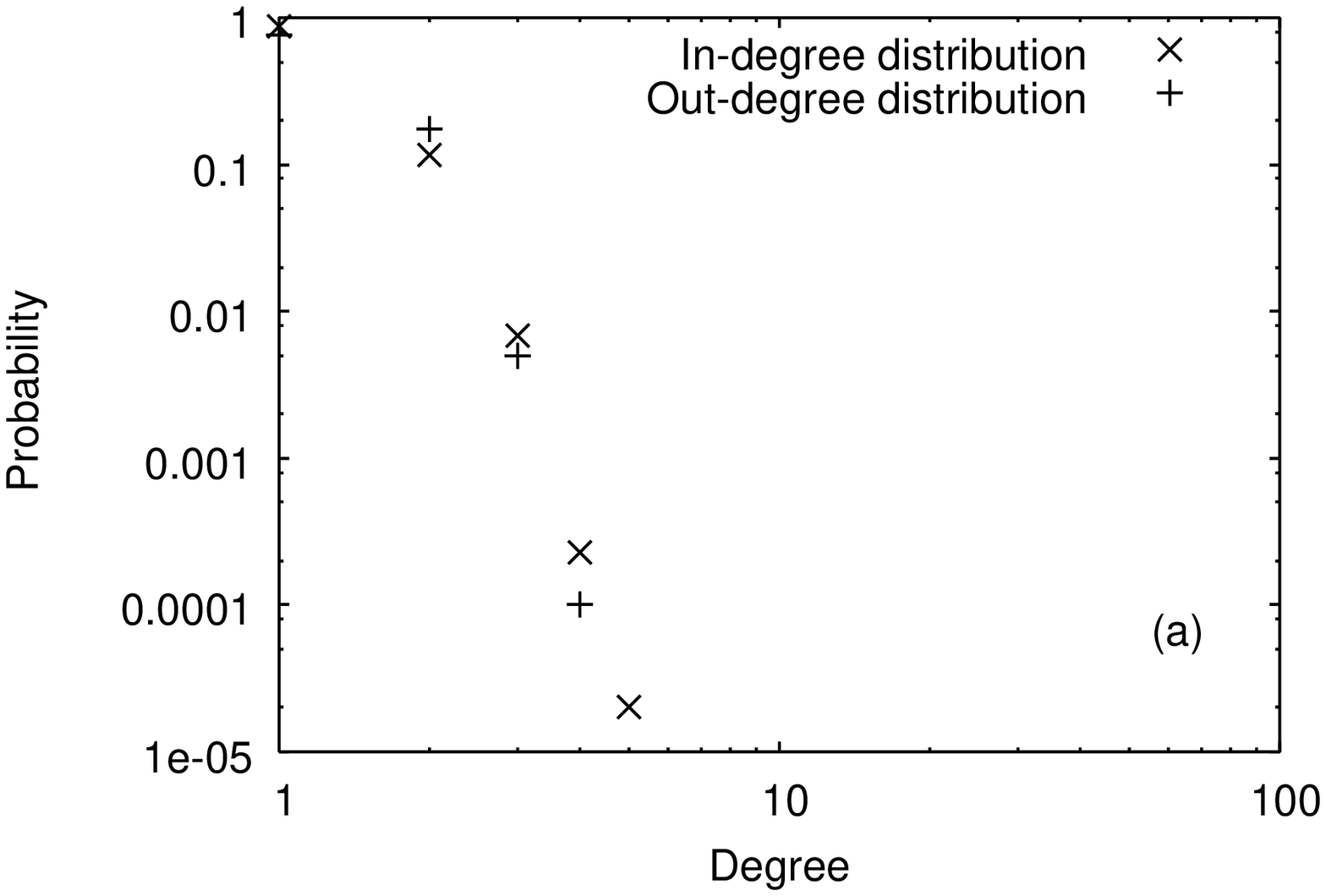}}&
\scalebox{0.32}{\includegraphics{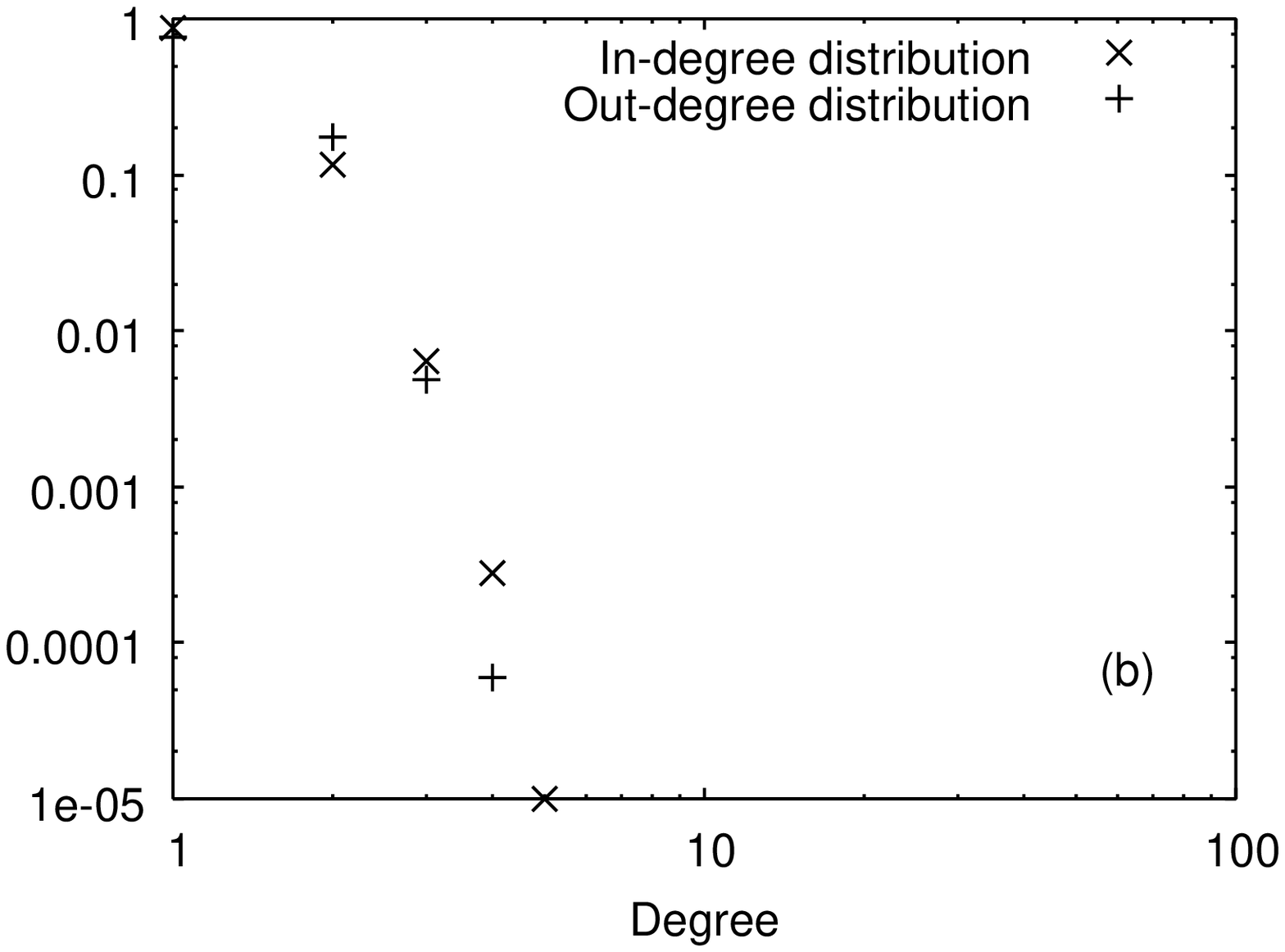}}
\end{tabular}
\caption{Average in- and out-degree distributions of $D_{0.2}^+(t)$ (a) and
$D_{0.8}^+(t)$ (b) for $t=10^4$ ($10^3$ independent runs).}
\label{distst28}
\vspace{0.45in}
\end{figure}

One possible exception to this conclusion may be the degree distributions of
$D^+_{0.2}(t)$ for $t=10^4$ (that is, well into the system's steady regime).
These two distributions are re-plotted from Figures~\ref{degdist2}(a) and (b) in
Figure~\ref{distst28}(a), where we can see a glimmer of the straight-line
behavior that characterizes power laws when plotted against doubly-logarithmic
scales. Confirmation that such a behavior is really present depends on
investigating substantially larger systems and also different values of $f$ to
see what trends exist. One example already appears in Figure~\ref{distst28}(b)
for the same system size as we have been investigating but with $f=0.8$, that
is, for a graph significantly sparser than $D_{0.2}^+(t)$. Seemingly, the
presence of power-law distributions is less likely for such a value of $f$.

\subsection*{Acknowledgments}

The authors acknowledge partial support from CNPq, CAPES, the PRONEX initiative
of Brazil's MCT under contract 41.96.0857.00, and a FAPERJ BBP grant. They also
thank Denise Carvalho and Alberto N\'obrega for their many insightful comments.


\begin{thebibliography}{28}
\expandafter\ifx\csname natexlab\endcsname\relax\def\natexlab#1{#1}\fi
\expandafter\ifx\csname url\endcsname\relax
  \def\url#1{\texttt{#1}}\fi
\expandafter\ifx\csname urlprefix\endcsname\relax\def\urlprefix{URL }\fi

\bibitem[{Abbas(2003)}]{a03}
Abbas, A.~K., 2003. Cellular and Molecular Immunology, 5th Edition. W. B.
  Saunders, Philadelphia, PA.

\bibitem[{Bak(1999)}]{b99}
Bak, P., 1999. How Nature Works: The Science of Self-Organized Criticality.
  Springer-Verlag, New York, NY.

\bibitem[{Bennett et~al.(1998)Bennett, Carbone, Karamalis, Flavell, Miller, and
  Heath}]{bckfmh98}
Bennett, S. R.~M., Carbone, F.~R., Karamalis, F., Flavell, R.~A., Miller,
  J.~F., Heath, W.~R., 1998. Help for cytotoxic {T}-cell responses is mediated
  by {CD}40 signalling. Nature 393, 478--480.

\bibitem[{Bernardes and dos Santos(1997)}]{bz97}
Bernardes, A.~T., dos Santos, R. M.~Z., 1997. Immune network at the edge of
  chaos. Journal of Theoretical Biology 186, 173--187.

\bibitem[{Bollob\'{a}s and Riordan(2003)}]{br03}
Bollob\'{a}s, B., Riordan, O.~M., 2003. Mathematical results on scale-free
  random graphs. In: Bornholdt, S., Schuster, H.~G. (Eds.), Handbook of Graphs
  and Networks. Wiley-VCH, Weinheim, Germany, pp. 1--34.

\bibitem[{Burnet(1957)}]{b57}
Burnet, F.~M., 1957. A modification of {J}erne's theory of antibody production
  using the concept of clonal selection. Australian Journal of Science 20,
  67--69.

\bibitem[{Burnet(1959)}]{b59}
Burnet, F.~M., 1959. The Clonal Selection Theory of Acquired Immunity.
  Cambridge University Press, Cambridge, UK.

\bibitem[{Coutinho(1995)}]{c95}
Coutinho, A., 1995. The network theory: 21 years later. Scandinavian Journal of
  Immunology 42, 3--8.

\bibitem[{De~Boer(1988)}]{db88}
De~Boer, R.~J., 1988. Symmetric idiotypic networks: connectance and switching,
  stability, and suppression. In: Perelson, A.~S. (Ed.), Theoretical
  Immunology: Part Two. Addison Wesley Longman, Redwood City, CA, pp. 265--289.

\bibitem[{De~Boer et~al.(1992)De~Boer, Segel, and Perelson}]{dbsp92}
De~Boer, R.~J., Segel, L.~A., Perelson, A.~S., 1992. Pattern formation in one-
  and two-dimensional shape-space models of the immune system. Journal of
  Theoretical Biology 155, 295--333.

\bibitem[{de~Castro and Timmis(2002)}]{dct02}
de~Castro, L.~N., Timmis, J., 2002. Artificial Immune Systems: A New
  Computational Intelligence Approach. Springer-Verlag, Heidelberg, Germany.

\bibitem[{Flores and Barbosa(2002)}]{fb02}
Flores, L.~E., Barbosa, V.~C., June 2002. A graph model for the evolution of
  specificity in immune systems. Tech. Rep. ES-582/02, COPPE, Federal
  University of Rio de Janeiro, Rio de Janeiro, Brazil.

\bibitem[{Forsdyke(1995)}]{f95}
Forsdyke, D.~R., 1995. The origins of the clonal selection theory of immunity:
  a case study for evaluation in science. The FASEB Journal 9, 164--166.

\bibitem[{Golub(1992)}]{g92}
Golub, E.~S., 1992. Is the function of the immune system only to protect? In:
  \cite{pw92}, pp. 15--26.

\bibitem[{Harada and Ikegami(2000)}]{hi00}
Harada, K., Ikegami, T., 2000. Evolution of specificity in an immune network.
  Journal of Theoretical Biology 203, 439--449.

\bibitem[{Holmberg et~al.(1989)Holmberg, Anderson, Carlsson, and
  Forsgen}]{hacf89}
Holmberg, D., Anderson, A., Carlsson, L., Forsgen, S., 1989. Establishment and
  functional implications of {B}-cell connectivity. Immunological Reviews 110,
  89--103.

\bibitem[{Jensen(1998)}]{j98}
Jensen, H.~J., 1998. Self-Organized Criticality: Emergent Complex Behavior in
  Physical and Biological Systems. Cambridge University Press, Cambridge, UK.

\bibitem[{Jerne(1974)}]{j74}
Jerne, N.~K., 1974. Towards a network theory of the immune system. Annales
  d'Immunologie C125, 373--389.

\bibitem[{Kleinstein and Seiden(2000)}]{ks00}
Kleinstein, S.~H., Seiden, P.~E., 2000. Simulating the immune system. Computing
  in Science and Engineering 2, 69--77.

\bibitem[{Maruyama et~al.(2000)Maruyama, Lam, and Rajewsky}]{mlr00}
Maruyama, M., Lam, K.~P., Rajewsky, K., 2000. Memory {B}-cell persistence is
  independent of persisting immunizing antigen. Nature 407, 636--642.

\bibitem[{Perelson and Oster(1979)}]{po79}
Perelson, A.~S., Oster, G.~F., 1979. The shape space model. Journal of
  Theoretical Biology 81, 645--670.

\bibitem[{Perelson and Weisbuch(1992)}]{pw92}
Perelson, A.~S., Weisbuch, G. (Eds.), 1992. Theoretical and Experimental
  Insights into Immunology. Vol.~66 of NATO ASI Series H: Cell Biology.
  Springer-Verlag, New York, NY.

\bibitem[{Perelson and Weisbuch(1997)}]{pw97}
Perelson, A.~S., Weisbuch, G., 1997. Immunology for physicists. Reviews of
  Modern Physics 69, 1219--1267.

\bibitem[{Ridge et~al.(1998)Ridge, di~Rosa, and Matzinger}]{rdrm98}
Ridge, J.~P., di~Rosa, F., Matzinger, P., 1998. A conditioned dendritic cell
  can be a temporal bridge between a {CD}4$^+$ {T}-helper and a {T}-killer
  cell. Nature 393, 474--478.

\bibitem[{Schoenberger et~al.(1998)Schoenberger, Toes, van~der Voort, Offringa,
  and Melief}]{stvdvom98}
Schoenberger, S.~P., Toes, R. E.~M., van~der Voort, E. I.~H., Offringa, R.,
  Melief, C. J.~M., 1998. {T}-cell help for cytotoxic {T} lymphocytes is
  mediated by {CD}40-{CD}40{L} interactions. Nature 393, 480--483.

\bibitem[{Segel and Cohen(2001)}]{sc01}
Segel, L.~A., Cohen, I.~R. (Eds.), 2001. Design Principles for the Immune
  System and Other Distributed Autonomous Systems. Oxford University Press, New
  York, NY.

\bibitem[{Silverstein(2002)}]{s02}
Silverstein, A.~M., 2002. The clonal selection theory: what it really is and
  why modern challenges are misplaced. Nature Immunology 3, 793--796.

\bibitem[{Stewart(1992)}]{s92}
Stewart, J., 1992. The immune system in an evolutionary perspective. In:
  \cite{pw92}, pp. 27--48.

\end{thebibliography}

\end{document}